\documentclass[aps,pre,twocolumn,eqsecnum,showpacs,floatfix]{revtex4}
\usepackage{graphicx}

\begin{document}

\title{Nonequilibrium critical dynamics of the relaxational models C and D}

\author{Vamsi K. \surname{Akkineni}}
\affiliation{Department of Physics, University of Illinois at Urbana-Champaign,
	     Urbana, IL 61801-3080 \,}
\email{akkineni@students.uiuc.edu}

\author{Uwe C. \surname{T\"auber}}
\affiliation{Department of Physics, Virginia Polytechnic Institute and State
             University, Blacksburg, VA 24061-0435 \,}
\email{tauber@vt.edu}

\date{\today}

\begin{abstract}
We investigate the critical dynamics of the $n$-component relaxational models C
and D which incorporate the coupling of a nonconserved and conserved order 
parameter ${\bf S}$, respectively, to the conserved energy density $\rho$, 
under nonequilibrium conditions by means of the dynamical renormalization 
group.
Detailed balance violations can be implemented isotropically by allowing for
different effective temperatures for the heat baths coupling to the slow modes.
In the case of model D with conserved order parameter, the energy density
fluctuations can be integrated out, leaving no trace of the nonequilibrium
perturbations in the asymptotic regime.
For model C with scalar order parameter, in equilibrium governed by strong
dynamic scaling ($z_S = z_\rho$), we find no genuine nonequilibrium fixed point
either.
The nonequilibrium critical dynamics of model C with $n=1$ thus follows the 
behavior of other systems with nonconserved order parameter wherein detailed 
balance becomes effectively restored at the phase transition.
For $n \geq 4$, the energy density generally decouples from the order 
parameter.
However, for $n=2$ and $n=3$, in the weak dynamic scaling regime 
($z_S \leq z_\rho$) entire lines of genuine nonequilibrium model C fixed points
emerge to one-loop order, which are characterized by continuously varying 
static and dynamic critical exponents.
Similarly, the nonequilibrium model C with spatially anisotropic noise and
$n < 4$ allows for continuously varying exponents, yet with strong dynamic 
scaling.
Subjecting model D to anisotropic nonequilibrium perturbations leads to 
genuinely different critical behavior with softening only in subsectors of 
momentum space and correspondingly anisotropic scaling exponents.
Similar to the two-temperature model B (randomly driven diffusive systems) the 
effective theory at criticality can be cast into an equilibrium model D 
dynamics, albeit incorporating long-range interactions of the uniaxial dipolar 
or ferroelastic type.
\end{abstract}



\maketitle

\section{Introduction}

Analytical studies of dynamic critical phenomena in the vicinity of a 
second-order phase transition usually rely on a coupled set of Langevin-type 
stochastic equations of motion for the relevant slow variables, namely the 
order parameter and hydrodynamic modes associated with conservation laws 
\cite{hohhal}.
Taking advantage of the separation of time scales induced by critical slowing 
down, all remaining microscopic degrees of freedom are reduced to additive
Gaussian white noise terms in this description.
In order to guarantee that the probability distribution for any configuration
converges to the canonical Gibbs function 
$\mathcal{P}_{\rm eq}(T) = Z(T)^{-1} \, \exp{(- \mathcal{H} / k_{\rm B} T})$ at
long times $t \to \infty$ (with the effective Hamiltonian $\mathcal{H}$ usually
taken to be the standard $\phi^4$ model), the second moments of the stochastic 
forces must be related to the relaxation rates via Einstein relations.
In addition, integrability conditions constrain the reversible force terms in
the nonlinear Langevin equations quite severely, for the associated probability
currents in the space of the slow variables must be divergence-free 
\cite{intcon}.
These two requirements also ensure the validity of the equilibrium 
fluctuation-dissipation theorem which relates the imaginary part of the dynamic
susceptibilities with the correlation functions.
As a consequence, the system's static behavior can be separated from its 
dynamic properties.

In isotropic systems, there are normally two independent static critical 
exponents, e.g., $\nu$ and $\eta$ which respectively characterize the 
divergence of the correlation length upon approaching the transition, 
$\xi \sim |\tau|^{-\nu}$, where $\tau \propto T - T_c$, and govern the 
power-law decay of the two-point correlation function at $T_c$, 
$C(|{\bf x}|) \sim |{\bf x}|^{-(d-2+\eta)}$ in $d$ spatial dimensions.
These become supplemented by dynamic exponents $z$ that describe the critical 
slowing down for the relevant modes, with characteristic relaxation times 
diverging as $t_{\rm c} \sim |\tau|^{- z \nu}$. 
At thermal equilibrium, the dynamic universality classes are well understood, 
and known to be distinguished by overall features of the dynamical system at 
hand.
In addition to the order parameter symmetry which essentially dictates the 
static critical exponents, the determining factors are if the order parameter 
itself represents a conserved quantity or not, the absence or presence of 
additional conservation laws, and the form of the reversible mode couplings  
between the generalized hydrodynamic variables, as again dictated by the 
symmetries of the problem \cite{hohhal}.

On the other hand, critical dynamics in systems far from thermal equilibrium is
not subject to the stringent limitations imposed by the detailed balance 
constraints, and in fact cannot even always be adequately captured through
coarse-grained stochastic equations of motion \cite{readif}.
Nevertheless, several important situations have been successfully modeled by
means of the Langevin formalism, two prominent examples being driven Ising
lattice gases, or more general driven diffusive systems \cite{drdifs}, and
nonequilibrium interface growth models, such as captured by the 
Kardar-Parisi-Zhang (KPZ) equation and its variants \cite{kapezt}.
Yet in nonequilibrium circumstances one has to invoke heuristic and/or 
phenomenological arguments for choosing the mathematical form of the noise 
correlations.
This must be done with appropriate care, however, since the structure of the 
stochastic forces may crucially impact the universal scaling behavior 
\cite{anikpz}.
It is thus of vital importance to elucidate the influence of different forms of
the assumed stochastic noise correlations on the long-distance and long-time
properties of any nonequilibrium Langevin system under investigation.

Naturally then, the following question arises for the Langevin models 
describing the equilibrium dynamical universality classes: 
What happens to their universal scaling behavior if the detailed balance 
conditions are violated ? 
Notice that since `static' properties cannot be decoupled from the dynamics in 
nonequilibrium steady states, this may include novel static critical behavior
as well in addition to perhaps modified values for the dynamic exponents $z$.
The simplest dynamical model just entails a purely relaxational kinetics for a
nonconserved order parameter with no coupling to other conserved quantities.
This defines model A in the (alphabetical) classification of 
Ref.~\cite{hohhal}.
Yet the model A universality class, as for example realized in the kinetic 
Ising model with Glauber spin flip dynamics, is known to be extremely robust 
against nonequilibrium perturbations \cite{fhaake,grins1}.
For the kinetic Ising model, this remains true even when the order parameter 
up-down symmetry is broken \cite{kevbea}.
Consider the most straightforward situation where the order parameter symmetry 
remains preserved, but the Einstein relation is not satisfied.
Since there is only a single stochastic equation of motion in this case, one 
can recover detailed balance through simple parameter rescaling which does not 
affect universal properties at the phase transition \cite{uwezol}.

Remarkably, the situation is markedly different for the purely diffusive
relaxational critical dynamics of model B with {\em conserved} order parameter
(e.g., the kinetic Ising model with Kawasaki spin exchange processes), but only
when subject to spatially {\em anisotropic} noise, say with stronger noise 
correlations in the thus defined longitudinal as compared to the complementary 
transverse sector in momentum space.
In this effective {\em two-temperature} or {\em randomly driven model B}, 
excitations in the transverse sector soften first, while the longitudinal 
directions remain noncritical \cite{bearoy,kevzol,beate2}.
This induces inherent anisotropic scaling at the critical point, of the same
form as those in driven lattice gases \cite{drdifs}.
Interestingly, the emerging long-wavelength dynamics in the critical regime can
be recast into an equilibrium model B, albeit with an effective Hamiltonian 
that incorporates long-range interactions of the uniaxial dipolar or 
ferroelastic type.
These reduce both the lower critical dimension, allowing long-range order 
already in one dimension, as well as the upper critical dimension to 
$d_c = 4 - d_\parallel$, where $d_\parallel$ denotes the dimension of the 
stiff longitudinal sector \cite{bearoy,kevzol}.
In Ref.~\cite{beate2}, the associated {\em four} independent critical exponents
($\nu$, $\eta$, $z$, and the anisotropy exponent $\Delta$) were computed to 
two-loop order in the $\epsilon$ expansion ($\epsilon = d_c - d$) by means of 
the dynamic renormalization group (RG), utilizing a path integral or dynamic 
field theory representation of the Langevin equation \cite{dynfun}.

The equilibrium dynamical models C and D (in the terminology of 
Ref.~\cite{hohhal}) still describe purely relaxational dynamics for either a 
nonconserved (model C) or conserved (model D) $n$-component order parameter 
field ${\bf S}$, which is however {\em statically} coupled to the conserved 
scalar energy density $\rho$ \cite{hahoma}.
Preserving the $O(n)$ order parameter symmetry, the lowest-order coupling
is $\propto \rho \, {\bf S}^2$.
As the energy density itself represents a noncritical variable entering the
Hamiltonian only quadratically, it can be integrated out exactly in the 
partition function $Z(T)$ and the generating function for static correlations.
This merely shifts the value of the fourth-order coupling $u$ for the order
parameter fluctuations, whence one recovers the static critical exponents of
the $O(n)$ model.

The coupling to the scalar diffusive mode $\rho$ may however alter the 
{\em dynamic} critical behavior.
To one-loop order, three distinct scaling regimes emerge for model C with
nonconserved order parameter, depending on the component number $n$ 
\cite{hahoma,bredom,folmos}:
(a) for Ising symmetry ($n = 1$), one finds `strong' scaling, i.e., 
$z_S = z_\rho = 2 + \alpha / \nu$, where $\alpha$ denotes the specific heat
critical exponent;
(b) the interval $2 \leq n < 4$, for which $\alpha > 0$, is characterized by
`weak' scaling with 
$z_S = 2 (1 + \alpha / n \nu) \leq z_\rho = 2 + \alpha / \nu$;
(c) for $n \geq 4$, where $\alpha \leq 0$, the Langevin equations for the 
${\bf S}$ and $\rho$ effectively decouple, leaving purely diffusive behavior
for the conserved mode, $z_\rho = 2$, and the model A dynamic critical exponent
for the order parameter, $z_S = 2 + c \, \eta$, with 
$c = 6 \ln \frac{4}{3} - 1 + O(\epsilon = 4-d)$.
To higher orders in perturbation theory, these three regimes essentially
persist (yet there appear additional distinctions with respect to the 
corrections to the leading scaling laws), but their boundaries become functions
of the spatial dimension $d$ as well as of $n$ \cite{bredom,folmos}.
For model D with conserved order parameter, the energy density always 
fluctuates faster in the critical region, rendering a strong-scaling regime
impossible.
The order parameter dynamics is thus not affected by the additional 
conservation law, and given by the model B dynamic critical exponent 
$z_S = 4 - \eta$.
For $\alpha > 0$, one finds again $z_\rho = 2 + \alpha / \nu$, whereas 
$z_\rho = 2$ in the decoupled case when $\alpha < 0$ \cite{hahoma}.

In this paper, we explore the effect of perturbations in the stochastic force
correlators that violate the equilibrium conditions on the critical dynamics of
the relaxational models C and D.
Specifically, we shall retain the $O(n)$ order parameter symmetry, but 
introduce different noise correlation strengths for the critical fluctuations 
and the conserved energy density, respectively, amounting to unequal effective 
heat bath temperatures $T_S$ and $T_\rho$.
We shall employ the dynamic RG to one-loop order, and search for novel
nonequilibrium fixed points of the ensuing RG flow equations.
In addition, we will investigate spatially anisotropic detailed balance 
violations.

The critical dynamics at structural phase transitions and of anisotropic 
antiferromagnets are usually listed as possible realizations of the model C 
universality class \cite{hohhal}. In the latter case, the nonequilibrium system
studied in this paper might be accessible experimentally if the effective
temperature of the conserved magnetization component(s) can be maintained at a
value different from that of the staggered magnetization which constitutes the 
nonconserved order parameter, perhaps through constant exposure to 
electromagnetic radiation.

This work supplements earlier research that focused on nonequilibrium 
perturbations for dynamic universality classes which are characterized by 
reversible mode couplings, as relevant for second-order phase transitions in 
magnetic systems \cite{uwezol} (models E, G, and J, respectively for the 
critical dynamics in planar ferromagnets, isotropic antiferromagnets, and 
Heisenberg ferromagnets) as well as in fluids \cite{jaiuwe} (model H for the 
liquid-gas transition critical point, or more generally in binary fluids, and 
model E for the normal- to superfluid phase transition).
Ref.~\cite{uwevjl} provides a concise summary of the results of these 
investigations (including a subset of this present work).

We finally remark that a recent study has addressed a {\em nonlocal} 
generalization of the equilibrium relaxational models that allows interpolating
between the scaling laws of models A, B, and C \cite{senbha}.
Intriguingly, also a `true model D' scaling regime emerges in this situation, 
where both the conservation laws for the order parameter and the energy density
remain relevant in the RG sense.

This article is organized as follows.
We start with a derivation of the Langevin equations of motion from the 
effective Hamiltonian for models C and D, and then provide a brief outline of 
the construction of field theory and the perturbation expansion based on the 
dynamic action (the Janssen-De Dominicis functional \cite{dynfun}). 
Prior to describing the results of the explicit perturbation expansion, the 
effect of the static nonlinear coupling of the order parameter to the conserved
energy density is gauged by integrating the conserved field out of the dynamic 
action. 
The implications of this, namely the reduction of the isotropic nonequilibrium 
model D to its equilibrium counterpart, is discussed. 
Subsequently, the full renormalization of the vertex functions (calculated to 
one-loop order) is detailed and the expressions for the resulting 
renormalization constants are presented. 
From these $Z$ factors we obtain the RG flow equations, and therefrom calculate
the static and dynamic critical exponents first in equilibrium, and then 
successively allowing for isotropic violation of detailed balance in both 
models D and C. 
The $Z$ factors are then adapted for the case of dynamical anisotropy in model 
C; its fixed point and critical behavior is explained. 
Finally, following earlier work on the two-temperature model B \cite{bearoy}, 
the anisotropic nonequilibrium model D is recast into an effective 
two-temperature model D with anisotropic scaling properties.
We conclude with a brief summary, putting our results into context with earlier
investigations.
An appendix lists the explicit expressions for the one-loop vertex functions.

\section{The relaxational models C and D}
\label{modelEqs} 

In this section, we outline the basic model equations for the nonequilibrium 
generalization of the relaxational models C and D. 
As introduced in Ref.~\cite{hahoma}, these models are characterized by a 
$n$-component vector order parameter 
${\bf S} \equiv \{S^\alpha\}, \alpha=1,\cdots,n$, coupled to a scalar conserved
field $\rho$. 
The effective Hamiltonian that describes their equilibrium static critical 
properties is the $O(n)$-symmetric $\phi^4$ Landau-Ginzburg-Wilson free energy 
in $d$ space dimensions, with additional terms for the noncritical conserved 
field and its coupling to the order parameter.
Preserving the $O(n)$ rotational invariance requires the lowest-order coupling 
to $\rho$ to be quadratic in ${\bf S}$.
For models C and D, the Hamiltonian thus reads
\begin{eqnarray}
&&\mathcal{H}[{\bf S},\rho] = \int \! d^dx \ \Biggl\{ \sum_{\alpha=1}^n 
\Biggl[ \frac{r}{2} \, S^{\alpha}({\bf x})^2 + \frac{1}{2} \, 
[\nabla S^{\alpha}({\bf x})]^2 \quad \label{hamilt} \\ 
&& + \frac{u}{4!} \sum_{\beta=1}^n S^{\alpha}({\bf x})^2 S^{\beta}({\bf x})^2
+ \frac{g}{2} \, \rho({\bf x}) S^{\alpha}({\bf x})^2 \Biggr] 
+ \frac{1}{2} \, \rho({\bf x})^2 \Biggr\} \ . \nonumber 
\end{eqnarray}
Here $r=(T-T_{c}^{o})/T_{c}^{o}$ denotes the relative distance from the 
mean-field critical temperature $T_{c}^{o}$, and we have rescaled the static
energy density correlation function, i.e., essentially the specific heat, to 
unity.
$u$ and $g$ represent the nonlinear interaction strengths.
The Hamiltonian (\ref{hamilt}) determines the equilibrium probability 
distribution for the fields $\{S^{\alpha}\}$ and $\rho$,
\begin{equation}
\mathcal{P}_{\rm eq}[{\bf S},\rho] = 
\frac{\exp(- \mathcal{H}[{\bf S},\rho] / k_B T)}{\int \mathcal{D}[{\bf S}] 
\mathcal{D}[\rho] \, \exp(- \mathcal{H}[{\bf S},\rho] / k_B T)} \ , 
\label{eqdist}
\end{equation}
and furthermore provides the starting point for the field-theoretic static 
renormalization group via a series expansion in the nonlinear couplings $u$ 
and $g$, which allows a systematic computation of the two independent static 
critical exponents $\eta$ and $\nu$ in powers of $\epsilon = 4 - d$. 
Notice that the energy density fluctuations $\rho$ enter the Hamiltonian 
(\ref{hamilt}) only linearly and quadratically, and can thus be readily 
integrated out. 
This merely results in a shift of the nonlinear coupling 
$u \to {\bar u} = u - 3 \, g^2$.
Provided the latter remains positive, this does not affect the RG fixed point
$u^*$, whence the static critical exponents are clearly those of the standard 
$O(n)$-symmetric $\phi^4$ model.

We now impose Langevin dynamics on the fluctuations of the order parameter and 
the conserved field to describe the relaxation of the system to equilibrium (at
which the stationarity conditions 
$\delta \mathcal{H}[{\bf S},\rho] / \delta S^\alpha = 0$ and 
$\delta \mathcal{H}[{\bf S},\rho] / \delta \rho = 0$ hold). 
The purely relaxational model C/D dynamics is then given by the equations of 
motion
\begin{eqnarray}
\frac{\partial S^{\alpha}({\bf x},t)}{\partial t} 
&=& - \lambda \, (i{\bf \nabla})^{a} \, 
\frac{\delta \mathcal{H}[{\bf S},\rho]}{\delta S^{\alpha}({\bf x},t)} 
+ \zeta^{\alpha}({\bf x},t) \ , \label{lgvneqn1} \\
\frac{\partial \rho({\bf x},t)}{\partial t} &=& D \, {\bf \nabla}^{2} \,
\frac{\delta \mathcal{H}[{\bf S},\rho]}{\delta \rho({\bf x},t)}
+ \eta({\bf x},t) \ . \label{lgvneqn2}
\end{eqnarray}
Here $\lambda$ and $D$ denote the relaxation coefficients of the order
parameter and energy density, respectively (i.e., $D$ is essentially the heat
conductivity). 
The distinction between models C and D is the value of the exponent $a$. 
For model C, $a=0$ corresponding to a {\em nonconserved} order parameter field,
while for model D, $a=2$, representing the diffusive relaxation of a 
{\em conserved} order parameter.
With the effective Hamiltonian (\ref{hamilt}), the equations of motion take the
specific form
\begin{eqnarray}
&&\frac{\partial S^{\alpha}({\bf x},t)}{\partial t} = - \lambda \, 
(i\nabla)^{a} \biggl[ (r-\nabla^2) S^\alpha({\bf x},t) + \frac{u}{6} \,
S^\alpha({\bf x},t) \nonumber \\
&&\ \times \sum_\beta S^\beta({\bf x},t)^2 + g \, \rho({\bf x},t) \, 
S^\alpha({\bf x},t) \biggr] + \zeta^\alpha({\bf x},t) \ , \label{lgvneqnI} \\
&&\frac{\partial \rho({\bf x},t)}{\partial t} = D \, \nabla^2 \biggl[
\rho({\bf x},t) + \frac{g}{2} \sum_\alpha S^\alpha({\bf x},t)^2 \biggr]
+ \eta({\bf x},t) \ . \nonumber \\ && \label{lgvneqnII}
\end{eqnarray}
Upon reinstating the specific heat, note that the linear part in 
Eq.~(\ref{lgvneqnII}) implies $t_c^{-1} \sim D \, |\tau|^\alpha \, q^2$.
Invoking dynamic scaling then suggests $t_c \sim |\tau|^{- 2 \nu - \alpha}$,
i.e., $z_\rho = 2 + \alpha / \nu$ for the dynamic exponent of the energy 
density.
We shall see that this relation in fact holds only for $n < 4$, or more
precisely, when $\alpha > 0$, for otherwise the two Langevin equations decouple
at criticality.

In the above stochastic equations of motion (\ref{lgvneqnI}) and 
(\ref{lgvneqnII}), $\zeta^\alpha$ and $\eta$ represent the stochastic forces 
(`noise') for the order parameter and the conserved field, respectively. 
We assume a Gaussian distribution for these fast variables with a vanishing 
temporal average, $\langle \zeta^\alpha \rangle = 0 = \langle \eta \rangle$. 
Their second moments then take the functional form
\begin{eqnarray}
\langle \zeta^\alpha({\bf x},t) \, \zeta^{\beta}({\bf x}^\prime,t^\prime) 
\rangle &=& 2 \widetilde{\lambda} \, (i\nabla)^{a} \, 
\delta ({\bf x}-{\bf x}^{\prime}) \, \delta(t-t^{\prime}) \, 
\delta^{\alpha\beta} \ , \nonumber \\ && \label{noiscorr1} \\
\langle \eta({\bf x},t) \, \eta({\bf x}^{\prime},t^{\prime}) \rangle 
&=& -2 \widetilde{D} \, \nabla^{2} \, \delta ({\bf x}-{\bf x}^{\prime}) \,
\delta (t-t^{\prime}) \, . \ \label{noiscorr2}
\end{eqnarray}

In thermal equilibrium, Einstein relations connect the relaxation coefficients 
with the corresponding noise strengths according to 
${\widetilde \lambda} = \lambda \, k_{\rm B} T$ and 
${\widetilde D} = D \, k_{\rm B} T$, where $T$ is the temperature of the 
heat bath in contact with the system. 
The detailed balance conditions implicit in these relations ensure that the 
system relaxes to the equilibrium probability distribution (\ref{eqdist}) in 
the limit $t \to \infty$. 
More generally, we can identify ${\widetilde \lambda}/ \lambda = k_{\rm B} T_S$
and ${\widetilde D} / D = k_{\rm B} T_\rho$ as the temperatures of the 
heat baths coupling to the order parameter and the conserved density, 
respectively, with their ratio given by 
\begin{equation}
\Theta = \frac{T_\rho}{T_S} = 
\frac{\widetilde D}{D} \ \frac{\lambda}{\widetilde \lambda} \ . \label{Theta}
\end{equation}
This new degree of freedom $\Theta$ describes the extent to which the 
equilibrium condition is violated; detailed balance clearly holds for 
$\Theta = 1$, while for $\Theta < 1$ energy flows from the order parameter heat
bath to the conserved density heat bath, and vice versa for $\Theta > 1$. 
Since we are interested in the behavior near the critical point 
$T_S \approx T_c$, $\Theta$ essentially measures the temperature of the 
conserved density heat bath $T_\rho$, in units of the critical temperature 
$T_c$. 
In the critical regime, we will be interested in calculating the dynamic 
exponents $z_S$ and $z_\rho$ that describes the critical slowing down for the 
order parameter and the conserved energy density. 
The one-loop RG flow equations for this case of {\em isotropic} detailed 
balance violation are derived in Sec.~\ref{isoCD}. 
The assumed functional form for the noise correlations (\ref{noiscorr1}),
(\ref{noiscorr2}) also enables us to impose a {\em spatially anisotropic} form 
of detailed balance violation, as described in Sec.~\ref{anistp}.

We close this general introduction with a brief outline of how a field theory 
representation can be constructed from general Langevin-type equations of the 
form
\begin{equation}
\frac{\partial \psi^{\alpha}({\bf x},t)}{\partial t} = 
K^{\alpha}[\{\psi^{\alpha}\}]({\bf x},t) + \zeta^{\alpha}({\bf x},t) \ ,
\label{geneom}
\end{equation}
with $\langle \zeta^{\alpha} \rangle = 0$, and the general noise correlations
\begin{equation}
\langle \zeta^{\alpha}({\bf x},t) \, \zeta^{\beta}({\bf x}^{\prime},t^{\prime})
\rangle = 2 L^{\alpha} \, \delta({\bf x}-{\bf x}^{\prime}) \, 
\delta (t-t^{\prime}) \, \delta^{\alpha\beta} \ . \label{gennoiscorr}
\end{equation}
This form of the white noise may be inferred from a Gaussian distribution for 
the stochastic forces 
\begin{equation}
W[\{\zeta^{\alpha}\}] \propto \exp\left[ -\frac{1}{4} \int \! d^{d}x \int \! dt
\sum_{\alpha} \zeta^{\alpha} (L^{\alpha})^{-1} \zeta^{\alpha} \right] \ .
\label{noisdist} 
\end{equation}
Eliminating $\zeta^{\alpha}$ via Eq.~(\ref{geneom}) then immediately yields the
desired probability distribution for the fields $\psi^{\alpha}$,
\begin{equation}
W[\{\zeta^{\alpha}\}] \mathcal{D}[\{\zeta^{\alpha}\}] = 
P[\{\psi^{\alpha}\}] \mathcal{D}[\{\psi^{\alpha}\}] \propto 
e^{G[\{\psi^{\alpha}\}]} \mathcal{D}[\{\psi^{\alpha}\}] \ ,
\label{flddistr}
\end{equation}
with the Onsager-Machlup functional
\begin{eqnarray}
G[\{\psi^{\alpha}\}] &=& -\frac{1}{4}\int \! d^{d}x \int \! dt \sum_{\alpha}
\left( \frac{\partial \psi^{\alpha}}{\partial t} - K^{\alpha}[\{\psi^\alpha\}] 
\right) \nonumber \\
&&\quad \times (L^\alpha)^{-1} \left( \frac{\partial \psi^\alpha}{\partial t} 
- K^{\alpha}[\{\psi^{\alpha}\}] \right) \ . \label{OMfnctl}
\end{eqnarray}
From this functional, one can already construct a perturbation expansion for 
the correlation functions of the fields $\psi^{\alpha}$; however, since the 
inverse of the Onsager coefficient $L^{\alpha}$ is singular for the conserved 
quantities, and furthermore high nonlinearities 
$\propto K^{\alpha}[\{\psi^{\alpha}\}]^{2}$ appear, it is convenient to 
introduce Martin-Siggia-Rose auxiliary fields via a Gaussian transformation to 
partially linearize the above functional. 
This leads to 
\begin{equation}
P[\{\psi^{\alpha}\}] \propto \int \! 
\mathcal{D}[\{i \widetilde{\psi}^{\alpha}\}] \, 
\exp(- {\mathcal A}[\{\widetilde{\psi}^{\alpha}\},\{\psi^{\alpha}\}])
\end{equation}
with the Janssen-De Dominicis functional \cite{dynfun}
\begin{eqnarray}
&&\mathcal{A}[\{\widetilde{\psi}^{\alpha}\},\{\psi^{\alpha}\}] = \int \! 
d^dx \! \int \! dt \ \sum_\alpha \Biggl[ - \widetilde{\psi}^\alpha L^\alpha 
\widetilde{\psi}^\alpha \nonumber \\ 
&&\qquad\qquad\qquad + \widetilde{\psi}^{\alpha} \left( 
\frac{\partial \psi^{\alpha}}{\partial t} - K^{\alpha}[\{\psi^{\alpha}\}] 
\right) \Biggr] \, . \label{JDfnctl1} 
\end{eqnarray}
Eq.~(\ref{JDfnctl1}) will provide the starting point for our discussion of the 
nonequilibrium dynamics of models C/D.
In Sec.~\ref{isoCD}, we will use the corresponding Janssen-De Dominicis 
functional for the construction of dynamic perturbation theory, and therefrom 
infer the one-loop RG flow equations, first in equilibrium and then with broken
detailed balance.
Subsequently in Sec.~\ref{anistp}, we will repeat the procedure for the models
with anisotropic detailed balance violation.

\section{The isotropic nonequilibrium models C and D}
\label{isoCD} 

In this section, we will study the nonequilibrium critical properties of the 
relaxational models C and D with isotropic violation of detailed balance. 
Along the way, we shall also recover the equilibrium critical exponents.
The field theory is constructed as outlined in the preceding 
Sec.~\ref{modelEqs}, and a perturbation series expansion in the relevant 
nonlinear couplings $\propto u$ and $g^2$ is developed for the one-particle 
irreducible vertex functions, explicitly here to one-loop order. 
The subsequent renormalization constitutes a straightforward generalization of 
the equilibrium renormalization scheme, see Ref.~\cite{bredom}. 
From the renormalization constants ($Z$ factors) that render the field theory 
finite in the ultraviolet (UV), we derive the RG flow functions which enter the
Callan--Symanzik equation. 
This partial differential equation describes the behavior of the correlation 
functions under scale transformations. 
In the vicinity of an RG fixed point, the theory becomes scale-invariant and 
the information from the UV behavior can be employed to access the physically 
interesting power laws governing the infrared (IR) regime at the critical point
($\tau \propto T-T_c \to 0$), for long wavelengths (${\bf q} \to 0$) and at low
frequencies ($\omega \to 0$).

\subsection{Dynamic field theory for models C and D}
\label{dftCD}

As a first step, we translate the Langevin equations (\ref{lgvneqnI}) and 
(\ref{lgvneqnII}), with the noise correlations (\ref{noiscorr1}) and 
(\ref{noiscorr2}), to a dynamic field theory \cite{dynfun,bredom}.
This results in a probability distribution for the dynamic fields ${\bf S}$ and
$\rho$
\begin{equation}
P[{\bf S},\rho] \propto \int \! {\cal D}[\{ i{\widetilde S}^\alpha \}] \int \!
{\cal D}[i{\widetilde \rho}] \ \exp (-\mathcal{A}[{\bf{\widetilde S}}, {\bf S},
{\widetilde \rho},\rho] \ , \label{dynprobdist}
\end{equation}
with the statistical weight given by the Janssen-De~Dominicis functional 
$\mathcal{A} = \mathcal{A}_{\rm har} + \mathcal{A}_{\rm rel} + 
\mathcal{A}_{\rm cd}$. 
The harmonic part, in terms of the original dynamic fields $S^\alpha$ and 
$\rho$, and the corresponding auxiliary fields ${\widetilde S}^\alpha$ and 
$\widetilde\rho$, reads
\begin{eqnarray}
&&\mathcal{A}_{\rm har}[{\bf{\widetilde S}},{\bf S},{\widetilde \rho},\rho] = 
\int \! d^dx \! \int \! dt \ \biggl( \sum_\alpha 
\widetilde{S}^\alpha({\bf x},t) \biggl[ 
\frac{\partial S^\alpha({\bf x},t)}{\partial t} \nonumber \\
&&\ + \lambda \, (i\nabla)^a \, (r-\nabla^2) \, S^\alpha({\bf x},t) - 
\widetilde{\lambda} \, (i\nabla)^a \, \widetilde{S}^\alpha({\bf x},t) \biggr]
\label{harJDfnctl} \\
&&\ + \widetilde{\rho}({\bf x},t) \biggl[ 
\frac{\partial \rho({\bf x},t)}{\partial t} - D \, \nabla^2 \, \rho({\bf x},t)
+ \widetilde{D} \, \nabla^2 \, \widetilde{\rho}({\bf x},t) \biggr] \biggr) \ ,
\nonumber 
\end{eqnarray}
while the static nonlinearity leads to a relaxation vertex,
\begin{eqnarray}
&&\mathcal{A}_{\rm rel}[{\bf{\widetilde S}},{\bf S}] = \frac{u}{6} \int \! d^dx
\! \int \! dt \ \sum_{\alpha,\beta} \widetilde{S}^\alpha({\bf x},t) \nonumber\\
&&\qquad\qquad\quad \times \biggl[ \lambda \, (i\nabla)^a \, 
S^\beta({\bf x},t)^2 \biggr] S^\alpha({\bf x},t) \ , \qquad \label{rlxJDfnctl}
\end{eqnarray}
and the coupling between the order parameter and the conserved density 
generates the model C/D vertices
\begin{eqnarray}
&&\mathcal{A}_{\rm cd}[{\bf{\widetilde S}},{\bf S},{\widetilde \rho},\rho] = g
\int \! d^dx \! \int \! dt \ \sum_\alpha \biggl[
\widetilde{S}^\alpha({\bf x},t) \, \lambda \, (i\nabla)^a \nonumber \\
&&\ \times \rho({\bf x},t) \, S^\alpha({\bf x},t) - \widetilde{\rho}({\bf x},t)
\, D \, \nabla^2 \, \frac{1}{2} \, S^\alpha({\bf x},t)^2 \biggr] \, . \quad 
\label{cdJDfnctl}
\end{eqnarray}

Before we proceed to develop the perturbation expansion based on the above
dynamic functional, we can try to gauge the relevance of the nonequilibrium 
parameter $\Theta$, as defined in Eq.~(\ref{Theta}), by integrating out the
conserved density $\rho$ from the action. 
Denoting those terms in the total dynamic action 
$\mathcal{A}[{\bf{\widetilde S}},{\bf S},{\widetilde \rho},\rho]$ that involve 
{\em only} the order parameter and the corresponding auxiliary fields as
$\mathcal{A}[{\bf{\widetilde S}},{\bf S}]$, and subtracting this part, we are,
in Fourier space, left with 
\begin{eqnarray}
&&\mathcal{A}[{\widetilde \rho},\rho] = 
\mathcal{A}[{\bf{\widetilde S}},{\bf S},{\widetilde \rho},\rho] - 
\mathcal{A}[{\bf{\widetilde S}},{\bf S}] \\
&&\ = \int \! \frac{d^dq}{(2\pi)^d} \! \int \! \frac{d\omega}{2\pi} \ \biggl\{
\widetilde{\rho}(-{\bf q},-\omega) \biggl[ (-i\omega + D \, {\bf q}^2) \,
\rho({\bf q},\omega) \nonumber \\
&&\qquad\qquad\qquad\qquad - \widetilde{D} \, {\bf q}^2 \, 
\widetilde{\rho}({\bf q},\omega) + D \, {\bf q}^2 \, \frac{g}{2} \, 
S^2({\bf q},\omega) \biggl] \nonumber \\ 
&&\qquad\qquad\qquad\qquad + \rho({\bf q},\omega) \, \lambda \, {\bf q}^a \, 
g \, [\widetilde{S} \cdot S](-{\bf q},-\omega) \biggr\} \ , \nonumber 
\label{JDrhotrms}
\end{eqnarray}
where we have introduced the composite operators
\begin{eqnarray}
&&S^2({\bf q},\omega) = \\
&&\qquad \int \! \frac{d^dp}{(2\pi)^d} \int \! \frac{d\nu}{2\pi} \ \sum_\alpha 
S^\alpha({\bf p},\nu) \, S^\alpha({\bf q}-{\bf p},\omega-\nu) \ , \nonumber 
\label{copss} \\ 
&&[\widetilde{S} \cdot S]({\bf q},\omega) = \\
&&\qquad \int \! \frac{d^dp}{(2\pi)^d} \int \! \frac{d\nu}{2\pi} \ \sum_\alpha 
\widetilde{S}^\alpha({\bf p},\nu) \, S^\alpha({\bf q}-{\bf p},\omega-\nu) 
\nonumber \label{copst}
\end{eqnarray}
as Fourier convolutions.

The path integral over the fields $\rho$ and $\widetilde \rho$ now takes the
form, in matrix notation,
\begin{eqnarray}
&&\int \! \mathcal{D}[i\widetilde\rho] \int \! \mathcal{D}[\rho] \, 
\exp(-\mathcal{A}[{\widetilde \rho},\rho]) \label{rhopathint1} \\
&&\quad = \prod_{{\bf q}, \omega} \int \! \mathcal{D}[ix_1] \mathcal{D}[x_2] 
\exp\left( - \frac{1}{2} \, x^T A \, x - b^T x \right) \ , \nonumber 
\end{eqnarray}
with the vectors
\begin{equation}
x = \left[ \begin{array}{c} \widetilde{\rho}({\bf q},\omega) \\ 
\rho({\bf q},\omega) \end{array} \right] , \quad 
b = \left[ \begin{array}{c} D \, {\bf q}^2 \, g \, S^2({\bf q},\omega) / 2 \\
\lambda \, {\bf q}^a \, g \, [\widetilde{S} \cdot S]({\bf q},\omega) 
\end{array} \right] , \label{xbdef}
\end{equation}
and the Hermitean matrix
\begin{equation}
A = \left[\begin{array}{cc} -2 \, \widetilde{D} \, {\bf q}^2 
& -i \omega + D \, {\bf q}^2 \\ i \omega + D \, {\bf q}^2 & 0 \end{array}
\right] = A^\dagger \ . \label{madef}
\end{equation}
After the linear transformation $y = x + A^{-1} \, b$, the integral 
(\ref{rhopathint1}) becomes
\begin{equation}
\int \! \mathcal{D}[iy_1] \mathcal{D}[y_2] \exp\left( -\frac{1}{2} \, y^T A \, 
y \right) \exp\left( \frac{1}{2} \, b^T A^{-1} \, b \right) \ ,
\label{rhopathint2}
\end{equation}
where the entries of the inverse matrix
\begin{equation}
A^{-1} = \left[\begin{array}{cc} 0 & (i \omega + D \, {\bf q}^2)^{-1} \\
(-i \omega + D \, {\bf q}^2)^{-1} & 2 \, \widetilde{D} \, {\bf q}^2 /
(\omega^2 + D^2 \, {\bf q}^4) \end{array} \right] \label{mainv}
\end{equation}
actually represent the propagators for the scalar conserved field $\rho$.
Upon performing the Gaussian integration over the $y$ fields, the $\rho$ and 
$\widetilde\rho$ fields are integrated out to yield the effective dynamic 
functional
\begin{eqnarray}
&&\mathcal{A}_{\rm eff}[{\bf{\widetilde S}},{\bf S}] =
\mathcal{A}[{\bf{\widetilde S}},{\bf S}] \label{effJDfnctl} \\ 
&&\quad + \int \! \frac{d^dq}{(2\pi)^d} \int \! \frac{d\omega}{2\pi} \ \lambda 
\, {\bf q}^a \, g^2 \, [\widetilde{S} \cdot S](-{\bf q},-\omega) \nonumber \\ 
&&\qquad \times \left[ \frac{S^2({\bf q},\omega) / 2}{1 - i \omega / D \, 
{\bf q}^2} + \frac{\lambda}{D} \ \frac{\widetilde{D} \, {\bf q}^a \,
[\widetilde{S} \cdot S]({\bf q},\omega) / D \, {\bf q}^2}{1 + (\omega / D \, 
{\bf q}^2)^2} \right] . \nonumber
\end{eqnarray}
We now define the parameter 
\begin{equation}
w = D / \lambda \ , \label{w}
\end{equation}
which essentially measures the ratio of relaxation times of the order parameter
and the conserved field, i.e., $w \sim \tau_S / \tau_\rho$. 
For model D ($a=2$) at criticality, the relaxation time of the conserved order 
parameter field is much longer (since 
$\partial S^\alpha / \partial t \sim {\bf q}^4$) compared to that of the 
conserved field (since $\partial \rho / \partial t \sim {\bf q}^2$), so that 
$w \to \infty$ as ${\bf q} \to 0$.
Hence the second term in the brackets in the above effective functional 
(which contains the ratio ${\widetilde D} / D$) vanishes asymptotically. 
Consequently, the nonequilibrium parameter $\Theta$ disappears from the 
effective field theory (\ref{effJDfnctl}) entirely, and a simple rescaling of 
the nonlinear couplings $u$ and $g$ reduces model D with {\em isotropic} 
detailed balance violation to its equilibrium counterpart. 
This remarkable result will be borne out in the explicit one-loop perturbation 
theory as well, see Sec.~\ref{isoD}.

\subsection{Perturbation theory and renormalization}
\label{pertthry}

\subsubsection{Elements of dynamic perturbation theory} 
We first detail the dynamic field theory for the case of isotropic detailed 
balance violation for both models C and D. 
The harmonic part (\ref{harJDfnctl}) defines the (bare) propagators of the 
field theory, while the perturbation expansion is performed in terms of the 
nonlinear vertices (\ref{rlxJDfnctl}) and (\ref{cdJDfnctl}). 
Note that the existence of the irreversible forces (\ref{cdJDfnctl}) does not
show up in dynamic mean-field theory (van Hove theory) at all, which is based 
on the harmonic action (\ref{harJDfnctl}) only.

We can now construct the perturbation expansion for all possible correlation 
functions of the dynamic and auxiliary fields, to be computed with the 
statistical weight 
$\exp(-\mathcal{A}[{\bf{\widetilde S}},{\bf S},{\widetilde \rho},\rho])$, as
well as for the associated vertex functions given by the one-particle 
irreducible Feynman diagrams. 
A straightforward scaling analysis yields that the upper critical dimension of
this model is $d_c = 4$ for the relaxational vertices (\ref{rlxJDfnctl}) and 
(\ref{cdJDfnctl}). 
Therefore, for $d \leq 4$ the perturbation theory will be IR-singular, and 
nontrivial critical exponents will result, whereas for $d \geq 4$ the 
perturbation theory contains UV divergences. 
In order to renormalize the field theory in the ultraviolet, it suffices to 
render all the nonvanishing two-, three-, and four-point functions finite by
introducing multiplicative renormalization constants (in addition to an 
additive renormalization that amounts to a fluctuation-induced shift of the 
critical temperature). 
This is achieved by demanding the renormalized vertex functions, or appropriate
momentum and frequency derivatives thereof, to be finite when the fluctuation 
integrals are taken at a conveniently chosen normalization point $\mu$, well 
outside the IR regime. 
Notice that $\mu$ defines an intrinsic momentum scale of the renormalized 
theory. 
The Callan--Symanzik equations can subsequently be used to explore the 
dependence of the {\em renormalized} vertex functions on $\mu$, and thereby 
obtain information on the scaling behavior of the dynamic correlation and 
response functions.

The Gaussian (zeroth-order) propagators 
\begin{eqnarray}
G^0_{\widetilde{S}^\alpha S^\beta}({\bf q},\omega) &=& 
\Gamma^0_{\widetilde{S}^\alpha S^\beta}(-{\bf q},-\omega)^{-1} \ , \\
G^0_{\widetilde{\rho} \rho}({\bf q},\omega) &=& 
\Gamma^0_{\widetilde{\rho} {\rho}}(-{\bf q},-\omega)^{-1}
\end{eqnarray}
and vertices which are the starting point for perturbation theory are
\begin{eqnarray}
\Gamma^0_{\widetilde{S}^\alpha S^\beta}({\bf q},\omega) &=& 
[i\omega + \lambda \, {\bf q}^a \, (r+{\bf q}^2)] \, \delta^{\alpha\beta} \, , 
\quad \\
\Gamma^0_{\widetilde{\rho} {\rho}}({\bf q},\omega) &=& 
i \omega + D \, {\bf q}^2 \ , \\
\Gamma^0_{\widetilde{S}^\alpha \widetilde{S}^\beta}({\bf q},\omega) &=& 
2 \, \widetilde{\lambda} \, {\bf q}^a \, \delta^{\alpha\beta} \ , \\
\Gamma^0_{\widetilde{\rho} \widetilde{\rho}}({\bf q},\omega) &=& 
2 \, \widetilde{D} \, {\bf q}^2 \ , \\
\Gamma^0_{\widetilde{S}^\alpha S^\beta \rho}({\bf q},\omega )&=& 
- \lambda \, {\bf q}^a \, g \, \delta^{\alpha\beta} , \\
\Gamma^0_{\widetilde{\rho} S^\alpha S^\beta}({\bf q},\omega) &=&
- \frac{1}{2} \, D \, {\bf q}^2 \, g \, \delta^{\alpha\beta} \ , \\
\Gamma^0_{\widetilde{S}^\alpha S^\alpha S^\beta S^\beta}({\bf q},\omega) &=& 
- \lambda \, {\bf q}^a \, \frac{u}{6} \ . 
\end{eqnarray}
In addition to the previously introduced ratio of relaxation times $w$ and the 
nonequilibrium parameter $\Theta$, we define for convenience the rescaled 
static couplings
\begin{equation}
\widetilde{u} = \frac{\widetilde{\lambda}}{\lambda} \ u \ , \quad
\widetilde{g}^2 = \frac{\widetilde{\lambda}}{\lambda} \ g^2 \ . \label{tug}
\end{equation}
Recall that $\Theta$ is a measure of the extent to which detailed balance is
violated, since for $\Theta = 1$ a straightforward rescaling of the couplings 
reduces the dynamic functional (\ref{harJDfnctl})--(\ref{cdJDfnctl}) to the
equilibrium case.

\subsubsection{Vertex function renormalization}
The explicit expressions for the relevant vertex 
functions to one-loop order in the perturbation expansion are given in 
appendix~\ref{appendix}. 
The ultraviolet-divergent derivatives of the two-, three-, and four-point 
vertex functions that require multiplicative renormalization are
$\partial_{q^a} \Gamma_{\widetilde{S}S}({\bf q},0) \mid_{q=0}$,
$\partial_{q^{2+a}} \Gamma_{\widetilde{S}S}({\bf q},0) \mid_{q=0}$,
$\partial_{\omega} \Gamma_{\widetilde{S}S}(0,\omega) \mid_{\omega=0}$,
$\partial_{q^2} \Gamma_{\widetilde{\rho}\rho}({\bf q},0) \mid_{q=0}$,
$\partial_{\omega} \Gamma_{\widetilde{\rho}\rho}(0,\omega) \mid_{\omega=0}$,
$\partial_{q^a} \Gamma_{\widetilde{S}\widetilde{S}}({\bf q},0) \mid_{q=0}$,
$\partial_{q^2}\Gamma_{\widetilde{\rho}\widetilde{\rho}}({\bf q},0)\mid_{q=0}$,
$\partial_{q^a} \Gamma_{\widetilde{S}S\rho}({\bf q},0) \mid_{q=0}$,
$\partial_{q^2} \Gamma_{\widetilde{\rho}SS}({\bf q},0) \mid_{q=0}$, and
$\partial_{q^a} \Gamma_{\widetilde{S}SSS}({\bf q},0)\mid_{q=0}$.
The quadratic divergence in the first of these will be taken care of by the 
$T_c$ shift $r_c$.
The remainder as well as all the other expressions are logarithmically
divergent at the upper critical dimension $d_c = 4$.
We thus require ten multiplicative renormalizations in all, which we take to
define the renormalized counterparts of the fields $\widetilde S$, $S$, 
$\widetilde \rho$, and $\rho$, and of the parameters $D$, $\widetilde D$,
$\lambda$, $\widetilde \lambda$, $u$, $g$, and $\tau=r-r_c$, which represents 
the temperature distance from the true critical temperature.  
The renormalized quantities are defined through
\begin{eqnarray}
&&S_R^\alpha = Z_S^{1/2} \, S^\alpha \, , \ 
\widetilde{S}_R^\alpha = Z_{\widetilde S}^{1/2} \, \widetilde{S}^\alpha \, , \\
&&\rho_R = Z_\rho^{1/2} \, \rho \, , \ 
\widetilde{\rho}_R = Z_{\widetilde \rho}^{1/2} \, \widetilde{\rho} \, , \\
&&\tau_R = Z_\tau \tau \, \mu^{-2} \, , \ \lambda_R = Z_\lambda \, \lambda \, ,
\ \widetilde{\lambda}_R = Z_{\widetilde \lambda} \widetilde{\lambda} \, , \\
&&D_R = Z_D\, D \, ,\ \widetilde{D}_R = Z_{\widetilde D} \, \widetilde{D}\, ,\\
&&u_R = Z_u u \, A_d \, \mu^{-\epsilon} \ , \
g_R^2 = Z_g g^2 \, A_d \, \mu^{-\epsilon} \, ,
\end{eqnarray}
where in standard notation
\begin{equation}
\epsilon = 4-d \ , \quad \textrm{and} \quad 
A_d = \frac{\Gamma(3-d/2)}{2^{d-1} \, \pi^{d/2}} \ . 
\end{equation}
The loop integrals are evaluated in the dimensional regularization scheme, and 
we choose the renormalized mass $\tau_R = 1$ as our normalization point 
(i.e., $\tau = \mu^2 Z_\tau^{-1} \approx \mu^2$, to lowest order).
Notice that as we have thus defined eleven renormalization constants ($Z$ 
factors), but actually only need ten multiplicative renormalizations, we shall 
have the freedom to choose, say, $\widetilde{\lambda}_R = \lambda_R$.
As $\widetilde{\lambda} = \lambda$ can be achieved through simple rescaling in
the unrenormalized theory, this implies the {\em choice} 
$Z_{\widetilde \lambda} = Z_\lambda$.
In addition, the structure of the perturbation series implies nontrivial 
additional {\em identities} between the renormalization constants, as we shall 
see below.

We now proceed to compute the renormalization factors by absorbing the UV 
divergences of the loop integrals following the minimal subtraction 
prescription. 
All subsequent explicit one-loop results are for the case of model C ($a = 0$).
However, as we have seen at the end of Sec.~\ref{dftCD}, the separation of 
relaxation time scales for the order parameter and the conserved density in 
model D leads to $w \to \infty$ in the asymptotic limit. 
It turns out that taking this limit for the model C results precisely yields 
the $Z$ factors for model D.
An independent calculation of the $Z$ factors from the model D vertex functions
confirms the validity of this simple limit procedure.

First, we employ the criticality condition 
$\chi(0,0)^{-1} = \Gamma_{\widetilde{S}S}(0,0) / \lambda = 0$ at the true
critical point $r=r_c$, and solve for the fluctuation-induced $T_c$ shift,
\begin{eqnarray}
r_c &=& - \frac{n+2}{6} \, (\widetilde{u} - 3 \widetilde{g}^2) \int_p \! 
\frac{1}{r_c+{\bf p}^2} \nonumber \\ 
&&- \widetilde{g}^2 \, \frac{1-\Theta}{1+w} \int_p \! 
\frac{1}{r_c/(1+w)+{\bf p}^2} \biggl] \ .
\end{eqnarray}
We may then reparametrize $\Gamma_{\widetilde{S}S}(0,0)$ in terms of 
$\tau=r-r_c$, which amounts to an additive renormalization, 
\begin{eqnarray}
&&\Gamma_{\widetilde{S}S}(0,0) = \lambda \, \tau \biggl [1 - \frac{n+2}{6} \,
(\widetilde{u} - 3 \widetilde{g}^2) \int_p \! 
\frac{1}{{\bf p}^2 \, (\tau+{\bf p}^2)} \nonumber \\
&&\qquad - \widetilde{g}^2 \, \frac{1-\Theta}{(1+w)^2} \int_p \! 
\frac{1}{{\bf p}^2 \, [\tau/(1+w)+{\bf p}^2]} \biggl] \, .
\end{eqnarray}
Writing this result in terms of renormalized quantities, and evaluating the
integrals at the normalization point $\tau = \mu^2$ in dimensional 
regularization, we obtain the following expression for the product of the $Z$ 
factors
\begin{eqnarray}
&&(Z_{\widetilde S} \, Z_S)^{1/2} \, Z_\lambda \, Z_\tau = \label{zprod1} \\
&&\quad 1 - \biggl[ \frac{n+2}{6} \, (\widetilde{u} - 3 \widetilde{g}^2) + 
\widetilde{g}^2 \, \frac{1-\Theta}{(1+w)^2} \biggr] 
\frac{A_d \mu^{-\epsilon}}{\epsilon} \ . \nonumber 
\end{eqnarray}
Next, expanding the integrands in the expression for 
$\Gamma_{\widetilde{S}S}({\bf q},\omega)$ to order ${\bf q}^2$ to obtain the 
renormalization factor for the relaxation rate $\lambda$, we find
\begin{eqnarray}
&&\frac{\partial}{\partial {\bf q}^2} \, \Gamma_{\widetilde{S}S}({\bf q},0) 
\Big\vert_{q=0} \\
&&\quad = \lambda \biggl[ 1 - \frac{\widetilde{g}^2}{4} \, \frac{1-\Theta}{1+w}
\int_p \! \frac{1}{[\tau / (1+w)+{\bf p}^2]^2} \nonumber \\
&&\qquad\quad + \frac{\widetilde{g}^2}{d} \, \frac{(1-\Theta)(1-w)^2}{(1+w)^3}
\int_p \! \frac{{\bf p}^2}{[\tau/(1+w)+{\bf p}^2]^3} \biggr] \ , \nonumber
\end{eqnarray}
whence
\begin{equation}
(Z_{\widetilde S} \, Z_S)^{1/2} \, Z_\lambda = 1 - \frac{\widetilde{g}^2}{4}
\, \frac{1-\Theta}{1+w} \biggl[ 1 - \frac{(1-w)^2}{(1+w)^2} \biggr] 
\frac{A_d \mu^{-\epsilon}}{\epsilon} \ . \label{zprod2} 
\end{equation}
Another product of $Z$ factors is obtained from
\begin{eqnarray}
&&\frac{\partial}{\partial (i\omega)} \, \Gamma_{\widetilde{S}S}(0,\omega)
\Big\vert_{\omega=0} \\
&&\quad = 1 + \frac{\widetilde{g}^2}{1+w} \int_p \! 
\frac{1}{(\tau+{\bf p}^2) \, [\tau/(1+w)+{\bf p}^2]} \nonumber \\
&&\qquad\quad - \widetilde{g}^2 \, \frac{1-\Theta}{(1+w)^2} \int_p \!
\frac{1}{[\tau/(1+w)+{\bf p}^2]^2} \ , \nonumber
\end{eqnarray}
which gives
\begin{equation}
(Z_{\widetilde S} \, Z_S)^{1/2} = 1 + \frac{\widetilde{g}^2}{1+w} \biggl[ 1 - 
\frac{1-\Theta}{1+w} \biggr] \frac{A_d \mu^{-\epsilon}}{\epsilon} \ . 
\label{zprod3}
\end{equation}

Now consider the two-point vertex function 
$\Gamma_{\widetilde{\rho}\rho}({\bf q},\omega)$ for the conserved density.
Because any loop diagram for this quantity necessarily involves a
$\widetilde{\rho} S S$ vertex for its outgoing leg, we see that taking 
${\bf q} \to 0$ results in 
$\Gamma_{\widetilde{\rho}\rho}(0,\omega) \equiv i\omega$ to {\em all} orders in
the perturbation expansion. 
As a consequence,
\begin{eqnarray}
Z_{\widetilde \rho} \, Z_\rho \equiv 1 \ . \label{zprod4}
\end{eqnarray}
Upon absorbing the logarithmic divergence of 
\begin{equation}
\Gamma_{\widetilde{\rho}\rho}({\bf q},0) = D \, {\bf q}^2 \biggl[ 1 - 
\frac{n}{2} \ \widetilde{g}^2 \int_p \! \frac{1}{(\tau+{\bf p}^2)^2} + O(q^4)
\biggr] 
\end{equation}
into $Z_D$, we arrive at
\begin{equation}
Z_D = 1 - \frac{n}{2} \ \widetilde{g}^2 \, 
\frac{A_d \mu^{-\epsilon}}{\epsilon} \ . \label{zprod5}
\end{equation}
The vertex function $\Gamma_{\widetilde{\rho}\widetilde{\rho}}({\bf q},\omega)$
is actually UV-finite to all orders. 
Again from the momentum dependence of the $\widetilde{\rho} S S$ vertex, 
$\partial_{q^2} \, \Gamma_{\widetilde{\rho}\widetilde{\rho}}({\bf q},0) 
\mid_{q=0} \equiv -2 \, \widetilde{D}$, whence
\begin{equation}
Z_{\widetilde\rho} \, Z_{\widetilde D} \equiv 1 \ , \label{zprod6}
\end{equation}
and with Eq.~(\ref{zprod4}) therefore 
\begin{equation}
Z_{\widetilde D} \equiv Z_\rho \ . \label{zprodI}
 \end{equation}
The remaining logarithmically divergent two-point function 
\begin{eqnarray}
&&\Gamma_{\widetilde{S}\widetilde{S}}(0,0) = \\
&&\quad -2 \, \widetilde{\lambda} \biggl[ 1 + \widetilde{g}^2 \, 
\frac{\Theta}{1+w} \int_p \! 
\frac{1}{(\tau+{\bf p}^2) \, [\tau/(1+w)+{\bf p}^2]} \biggr] \nonumber
\end{eqnarray}
yields the relation
\begin{equation}
Z_{\widetilde S} \, Z_{\widetilde \lambda} = 1 + \widetilde{g}^2 \, 
\frac{\Theta}{1+w} \, \frac{A_d \mu^{-\epsilon}}{\epsilon} \ . \label{zprod7}
\end{equation}

At last, from the rather lengthy one-loop results 
(\ref{threept1})--(\ref{fourpt1}) for the three- and four-point vertex 
functions we deduce, respectively,
\begin{eqnarray}
&&Z_S \, (Z_{\widetilde \rho} \, Z_g)^{1/2} \, Z_D = \label{zprod9} \\
&&\quad 1 - \frac{n+2}{6} \, \widetilde{u} \, \frac{A_d \mu^{-\epsilon}}
{\epsilon} + \widetilde{g}^2 \biggl[ 1 - \frac{1-\Theta}{1+w} \biggr] 
\frac{A_d \mu^{-\epsilon}}{\epsilon} \ , \nonumber \\
&&(Z_{\widetilde S} \, Z_S \, Z_\rho \, Z_g)^{1/2} \, Z_\lambda = 
\label{zprod10} \\
&&\quad 1 - \frac{n+2}{6} \, \widetilde{u} \, \frac{A_d \mu^{-\epsilon}}
{\epsilon} + \widetilde{g}^2 \biggl[ 1 - \frac{1-\Theta}{(1+w)^2} \biggr] 
\frac{A_d \mu^{-\epsilon}}{\epsilon} \ , \nonumber \\
&&(Z_{\widetilde S}\, Z_S)^{1/2} \, Z_S \, Z_\lambda \, Z_u = 1 - 
\frac{n+8}{6} \, \widetilde{u} \, \frac{A_d \mu^{-\epsilon}}{\epsilon} 
\nonumber \\
&&\qquad\qquad + 6 \, \widetilde{g}^2 \biggl[ 1 - \frac{(1-\Theta) \, (2+w)}
{2 \, (1+w)^2} \biggr] \frac{A_d \mu^{-\epsilon}}{\epsilon} \nonumber \\
&&\qquad\qquad\qquad - \frac{6 \, \widetilde{g}^4}{\widetilde{u}} \biggl[ 1 - 
\frac{1-\Theta}{1+w} \biggr] \frac{A_d \mu^{-\epsilon}}{\epsilon} \ . 
\label{zprod8}
\end{eqnarray}

Upon factoring from the above $Z$ factor products 
(\ref{zprod1}), (\ref{zprod2}), (\ref{zprod3}), (\ref{zprod5}), and
(\ref{zprod9})--(\ref{zprod8}), the following results are obtained:
\begin{eqnarray}
&&\qquad Z_\tau = 1 - \frac{n+2}{6} \, \widetilde{u} \frac{A_d \mu^{-\epsilon}}
{\epsilon} \nonumber \\
&&\qquad\qquad\quad + \widetilde{g}^2 \biggl[ \frac{n+2}{2} - \frac{1-\Theta}
{(1+w)^3} \biggr] \frac{A_d \mu^{-\epsilon}}{\epsilon} \ , \label{zprodVII} \\
&&\qquad Z_\lambda = 1 - \frac{\widetilde{g}^2}{1+w} \biggl[ 1 - 
\frac{1-\Theta}{(1+w)^2} \biggr] \frac{A_d \mu^{-\epsilon}}{\epsilon} \ ,
\label{zprodIV} \\
&&Z_S \, Z_\rho^{-1} = 1 + \widetilde{g}^2 \biggl[ \frac{n}{2} - (1-\Theta) 
\, \frac{w \, (2+w)}{(1+w)^3} \biggr] \frac{A_d \mu^{-\epsilon}}{\epsilon} \ , 
\nonumber \\ && \label{zprodX} \\
&&Z_S \, (Z_\rho^{-1} \, Z_g)^{1/2} = 1 - \frac{n+2}{6} \, \widetilde{u} \,
\frac{A_d\mu^{-\epsilon}}{\epsilon} \nonumber \\
&&\qquad\qquad\qquad + \widetilde{g}^2 \biggl[ \frac{n+2}{2} - \frac{1-\Theta}
{1+w} \biggr] \frac{A_d \mu^{-\epsilon}}{\epsilon} \ , \label{zprodVIII} \\
&&(Z_\rho \, Z_g)^{1/2} = 1 - \frac{n+2}{6} \, \widetilde{u} \,
\frac{A_d \mu^{-\epsilon}}{\epsilon} \nonumber \\
&&\qquad\qquad\qquad + \widetilde{g}^2 \biggl[ 1 - \frac{1-\Theta}{(1+w)^3} 
\biggr] \frac{A_d \mu^{-\epsilon}}{\epsilon} \ , \label{zprodIX} \\
&&Z_S \, Z_g = 1 - \frac{n+2}{3} \, \widetilde{u} \, 
\frac{A_d \mu^{-\epsilon}}{\epsilon} \label{zprodXI} \\
&&\qquad\ + \widetilde{g}^2 \biggl[ \frac{n+4}{2} - (1-\Theta) \, 
\frac{2+2w+w^2}{(1+w)^3} \biggr] \frac{A_d \mu^{-\epsilon}}{\epsilon} \ , 
\nonumber \\
&&Z_S \, Z_u = 1 - \frac{n+8}{6} \, \widetilde{u} \, 
\frac{A_d \mu^{-\epsilon}}{\epsilon} \nonumber \\
&&\quad + \widetilde{g}^2 \biggl[ 6 - \frac{1-\Theta}{1+w} \biggl(
\frac{3(2+w)}{1+w} - \frac{w}{(1+w)^2} \biggr) \biggr] 
\frac{A_d \mu^{-\epsilon}}{\epsilon} \nonumber \\
&&\qquad\qquad\qquad - \frac{6 \, \widetilde{g}^4}{\widetilde{u}} \biggl[ 1 - 
\frac{1-\Theta}{1+w} \biggr] \frac{A_d \mu^{-\epsilon}}{\epsilon} \ , 
\label{zprodVI}
\end{eqnarray}
supplementing Eqs.~(\ref{zprod3}), (\ref{zprod4}), (\ref{zprod5}),
(\ref{zprod6}), (\ref{zprodI}), and (\ref{zprod7}).

For model D ($a=2$), the ratio of relaxation times $w$ constitutes a relevant
parameter in the RG sense, whence $w \to \infty$ asymptotically.
Obviously, this leads to marked simplifications in the above expressions for 
the renormalization constants.
In addition, as a consequence of the order parameter conservation law and the
ensuing momentum dependence of the vertices, we have 
$\Gamma_{\widetilde{S}S}(0,\omega) \equiv i\omega$ and 
$\partial_{q^2} \, \Gamma_{\widetilde{S}\widetilde{S}}({\bf q},0) \mid_{q=0} 
\equiv -2 \, \widetilde{\lambda}$ to all orders in perturbation theory, which
implies the relations $Z_{\widetilde S} \, Z_S \equiv 1$, 
$Z_{\widetilde S} \, Z_{\widetilde \lambda} \equiv 1$, and thus 
$Z_{\widetilde \lambda} \equiv Z_S$, which also follow to one-loop order from 
Eqs.~(\ref{zprod3}) and (\ref{zprod7}), respectively.

\subsubsection{Callan--Symanzik and RG flow equations}
\label{RGeqns} 
By means of the above renormalization constants, we can now write down the 
Callan--Symanzik RG equations for the vertex functions and the dynamic 
susceptibilities, which describe the dependence on the renormalization scale 
$\mu$, and thus on the renormalized couplings.
These RG equations connect the asymptotic theory, where the IR singularities 
become manifest, with a region in parameter space where the loop integrals are 
finite and ordinary `naive' perturbation expansion is applicable.
They follow from the observation that the `bare' vertex functions do not depend
on the renormalization scale $\mu$,
\begin{equation}
\mu \, \frac{d}{d\mu} \bigg\vert_0 
\Gamma^{{\widetilde S}^m {\widetilde \rho}^n S^r \rho^s}(\{ {\bf q},\omega \} ;
\{ a \}) = 0 \ , \label{rngreq}
\end{equation}
where $\{ a \}$ represents the parameter set $u$, $g^2$, ${\widetilde D}$, 
$D$, ${\widetilde \lambda}$, $\lambda$, and $\tau$. 
Replacing the bare parameters and fields in Eq.~(\ref{rngreq}) with the 
renormalized ones, we find the following partial differential equations for the
renormalized vertex functions:
\begin{eqnarray}
&&\biggl[ \mu \, \frac{\partial}{\partial \mu} + \sum_{\{ a_R \}} \gamma_a \,
a_R \, \frac{\partial}{\partial a_R} + \frac{m}{2} \, \gamma_{\widetilde S} + 
\frac{n}{2} \, \gamma_{\widetilde \rho} + \frac{r}{2} \, \gamma_S \nonumber \\
&&\qquad + \frac{s}{2} \, \gamma_\rho \biggr] 
\Gamma_R^{{\widetilde S}^m {\widetilde \rho}^n S^r \rho^s}(\{{\bf q},\omega \};
\{ a_R \}) = 0 \ . \label{calsym}
\end{eqnarray}
Here, we have introduced Wilson's flow functions
\begin{eqnarray}
&&\gamma_{\widetilde S} = \mu \, \frac{\partial}{\partial \mu} \bigg\vert_0
\ln Z_{\widetilde S} \, , \ 
\gamma_S = \mu \, \frac{\partial}{\partial \mu} \bigg\vert_0 \ln Z_S \ ,
\label{flowfct1} \\
&&\gamma_{\widetilde \rho} = \mu \, \frac{\partial}{\partial \mu} \bigg\vert_0
\ln Z_{\widetilde \rho} \, , \
\gamma_\rho = \mu \, \frac{\partial}{\partial \mu} \bigg\vert_0 \ln Z_\rho
\label{flowfct2}
\end{eqnarray}
for the fields and
\begin{equation}
\gamma_a = \mu \, \frac{\partial}{\partial \mu} \bigg\vert_0 \ln \frac{a_R}{a}
\label{flowfct3}
\end{equation}
for the different parameters (the subscript `0' indicates that the renormalized
fields and parameters are to be expressed in terms of their bare counterparts 
prior to taking the derivatives with respect to the momentum scale $\mu$).

The Callan--Symanzik equations (\ref{calsym}) are solved via the method of 
characteristics, introducing $\hat\mu(\ell) = \mu \ell$, where $\ell$ is a real
continuous parameter.
This defines running couplings as the solutions to the first-order differential
RG flow equations
\begin{equation}
\ell \, \frac{d {\hat a}(\ell)}{d\ell} = \gamma_a(\ell) \, {\hat a}(\ell) \ , 
\quad \textrm{with} \quad {\hat a}(1) = a_R \ . \label{rgflow}
\end{equation}
The solutions of the partial differential equations (\ref{calsym}) then read
\begin{eqnarray}
&&\Gamma_R^{{\widetilde S}^m {\widetilde \rho}^n S^r \rho^s}(\mu,
\{ {\bf q},\omega \};\{ a_R \}) = \label{solcsy} \\
&&\quad \exp \biggl( \int_1^\ell \! \frac{d \ell'}{\ell'} \biggl[ \frac{m}{2} 
\, \gamma_{\widetilde S}(\ell') + \frac{r}{2} \, \gamma_S(\ell') 
+ \frac{n}{2} \, \gamma_{\widetilde \rho}(\ell') \nonumber \\
&&\qquad\qquad + \frac{s}{2} \, \gamma_\rho(\ell') \biggr] \biggr) \,
\hat\Gamma_R^{{\widetilde S}^m {\widetilde \rho}^n S^r \rho^s}(\mu \ell, 
\{ {\bf q},\omega \};\{ \hat a(\ell) \}) \ . \nonumber 
\end{eqnarray}

\subsection{Model C and D equilibrium critical exponents}
\label{eqlbexp} 

To begin the analysis of the RG flow equations, we recover the critical 
exponents for the equilibrium models C and D (see the original 
Ref.~\cite{hahoma} and Ref.~\cite{bredom} for the corresponding field theory; 
the two-loop analysis was recently clarified in Ref.~\cite{folmos}).
Upon removing the effects of the nonequilibrium perturbation by setting 
$\widetilde{\lambda} = \lambda$ and $\widetilde{D} = D$ (we may put 
$k_{\rm B} T \approx k_{\rm B} T_c = 1$), whence $\Theta=1$, $\widetilde{u}=u$,
and $\widetilde{g}^2=g^2$ in the previous expressions for the $Z$ factors, we 
obtain the renormalization constants for the equilibrium case to one-loop 
order:
\begin{eqnarray}
&&Z_\rho = Z_{\widetilde\rho}^{-1} = Z_{D} = 1 - \frac{n}{2} \
\frac{g^2 \, A_d \mu^{-\epsilon}}{\epsilon} \ , \label{zfacteq1} \\
&&Z_S = 1 \ , \label{zfacteq2} \\
&&Z_{\widetilde S}^{-1/2} = Z_\lambda = 1 - \frac{1}{1+w} \
\frac{g^2 \, A_d \mu^{-\epsilon}}{\epsilon} \ , \label{zfacteq3} \\
&&Z_\tau = 1 - \frac{n+2}{6} \ \frac{u \, A_d \mu^{-\epsilon}}{\epsilon} + 
\frac{n+2}{2} \ \frac{g^2 \, A_d \mu^{-\epsilon}}{\epsilon} \ , 
\label{zfacteq4} \\
&&Z_g = 1 - \frac{n+2}{3} \ \frac{u \, A_d \mu^{-\epsilon}}{\epsilon} +
\frac{n+4}{2} \ \frac{g^2 \, A_d \mu^{-\epsilon}}{\epsilon} \ , 
\label{zfacteq5} \\
&&Z_u = 1 - \frac{n+8}{6} \ \frac{u \, A_d \mu^{-\epsilon}}{\epsilon} + 6
\left( 1 - \frac{g^2}{u} \right) \frac{g^2 \, A_d \mu^{-\epsilon}}{\epsilon} 
\ . \nonumber \\ && \label{zfacteq6}
\end{eqnarray}
In this equilibrium system, there exists a fluctuation-dissipation relation 
that relates the imaginary part of the dynamic order parameter susceptibility 
$\chi({\bf q},\omega)$ to the Fourier transform of the dynamic correlation 
function $C({\bf x}-{\bf x}',t-t') \, \delta^{\alpha\beta} = 
\langle S^\alpha({\bf x},t) \, S^\beta({\bf x}',t') \rangle$:
\begin{equation}
C({\bf q},\omega) = \frac{2 k_{\rm B} T}{\omega} \ \textup{Im} \, 
\chi({\bf q},\omega) \ .
\end{equation}
These quantities are connected to the two-point vertex functions via
$C({\bf q},\omega) = - \Gamma_{{\widetilde S}{\widetilde S}}({\bf q},\omega) 
/ |\Gamma_{{\widetilde S}S}({\bf q},\omega)|^2$, and $\chi({\bf q},\omega) = 
\lambda {\bf q}^a / \Gamma_{{\widetilde S}S}(-{\bf q},-\omega)$; thus the 
fluctuation-dissipation theorem results in the following relation between the 
two-point vertex functions
\begin{eqnarray}
\Gamma_{{\widetilde S}{\widetilde S}}({\bf q},\omega) = 
\frac{2 \widetilde{\lambda}}{\omega} \ \textup{Im} \, 
\Gamma_{{\widetilde S}S}({\bf q},\omega) \ . \label{fdtver}
\end{eqnarray}
The same identity must hold in the renormalized theory.
Consequently,
\begin{equation}
Z_\lambda \equiv \left( Z_S / Z_{\widetilde S} \right)^{1/2}\ , \label{zprodDS}
\end{equation}
and in the same manner for the conserved field, 
\begin{equation}
Z_D \equiv \left( Z_\rho / Z_{\widetilde \rho} \right)^{1/2}\ . \label{zprodDD}
\end{equation}
Both relations are indeed fulfilled by the above explicit one-loop results.
Through taking logarithmic derivatives with respect to the normalization scale
$\mu$, one finds that the equilibrium fluctuation-dissipation theorem implies
\begin{equation}
2 \, \gamma_\lambda \equiv \gamma_S - \gamma_{\widetilde S} \quad \textup{and} 
\quad 2 \, \gamma_D \equiv \gamma_\rho - \gamma_{\widetilde \rho} \ .
\label{fdtflow}
\end{equation}

The explicit RG flow functions derived from the one-loop renormalization 
constants become
\begin{eqnarray}
&&\gamma_S = 0 \, , \ \gamma_\lambda = - \, \frac{\gamma_{\widetilde S}}{2} 
= \frac{g_R^2}{1+w_R} \ , \label{gammaeq1} \\
&&\gamma_\rho = - \gamma_{\widetilde \rho} = \gamma_D = \frac{n}{2} \ g_R^2 \ ,
\label{gammaeq2} \\
&&\gamma_\tau = - 2 + \frac{n+2}{6} \ {\bar u}_R \ . \label{gammaeq4}
\end{eqnarray}
In the result for the single nontrivial static RG flow function 
(\ref{gammaeq4}) to one-loop order, ${\bar u}_R = u_R - 3 \, g_R^2$ represents
the shifted coupling that also results from directly integrating out the scalar
density $\rho$, see Sec.~\ref{modelEqs}.
Indeed, from Eqs.~(\ref{zfacteq5}) and (\ref{zfacteq6}) we infer its $Z$ factor
\begin{equation}
Z_{\bar u} = 1 - \frac{n+8}{6} \ 
\frac{{\bar u} \, A_d \mu^{-\epsilon}}{\epsilon} \ , \label{zfacteq7}
\end{equation}
which along with Eq.~(\ref{gammaeq4}) is just the standard one-loop result for 
the $O(n)$-symmetric $\phi^4$ theory.
Note that for model D, taking $w_R \to\infty$ in Eq.~(\ref{gammaeq1}) yields 
$\gamma_\lambda = 0$. 

We are now in a position to study the scaling behavior in the vicinity of the 
various RG fixed points which are given by the zeros of the RG $\beta$ 
functions
\begin{equation}
\beta_a = \gamma_a a_R = \mu \, \frac{\partial}{\partial \mu} \bigg\vert_0 a_R
\end{equation}
for the nonlinear couplings $\bar u$ and $g^2$ as well as the relaxation rate
ratio $w$, $\beta_w = w_R \, (\gamma_D - \gamma_\lambda)$.
By means of Eqs.~(\ref{zfacteq1}), (\ref{zfacteq3}), (\ref{zfacteq5}), and 
(\ref{zfacteq7}), we find
\begin{eqnarray}
\beta_{\bar u} &=& {\bar u}_R \left[ -\epsilon + \frac{n+8}{6} \ {\bar u}_R
\right] \ , \label{eqbetu} \\
\beta_g &=& g_R^2 \left[ -\epsilon + \frac{n+2}{3} \ {\bar u}_R + 
\frac{n}{2} \ g_R^2 \right] \ , \label{eqbetg} \\
\beta_w &=& w_R \, g_R^2 \left[ \frac{n}{2} - \frac{1}{1+w_R} \right] \ . 
\end{eqnarray}
For model C, the flow function $\beta_w$ yields three fixed points, provided 
${g^\ast}^2 > 0$, namely $w_0^\ast = 0$, $w_C^\ast = (2/n)-1$ (which is 
positive for $0 < n < 2$), and $w_D^\ast = \infty$.
Stability requires that 
\begin{equation}
\frac{\partial \beta_w}{\partial w_R} = g_R^2 \left[ \frac{n}{2} - 
\frac{1}{(1+w_R)^2} \right] 
\end{equation}
be positive at the fixed point.
Consequently, for $n=1$ we find that $w_C^\ast = 1$ is stable, whereas 
$w_0^\ast = 0$ for $n \geq 2$.
Recall that $w_D^\ast = \infty$ corresponds to model D; this fixed point is
unstable in model C for all values of $n$.

For the static coupling $\bar u_R$ we find the following zeros of
$\beta_{\bar u}$: the Gaussian fixed point $u^{\ast}_0 = 0$, and the 
Heisenberg fixed point $u^{\ast}_H = 6 \, \epsilon / (n+8)$. 
Inserting these in turn into the flow function $\beta_g$, we obtain: 
${g^\ast_0}^2 = 0$ and ${g^\ast_1}^2 = 2 \, \epsilon / n$ corresponding to 
$u^{\ast}_0 = 0$; ${g^\ast_0}^2=0$ and 
${g^\ast_C}^2 =  2 \, (4-n) \, \epsilon / n \, (n+8)$ corresponding to
$u^{\ast}_H$. 
The stability of these four fixed points in the $({\bar u}_R,g_R^2)$ plane 
depends on the spatial dimension $d$ and the number of order parameter 
components $n$.
Checking for positivity of the eigenvalues of the stability matrix with the
entries $\partial \beta_{\bar u} / \partial {\bar u}_R = - \epsilon + (n+8) \, 
{\bar u}_R / 3$, $\partial \beta_{\bar u} / \partial g^2_R = 0$, 
$\partial \beta_g / \partial {\bar u}_R = (n+2) \, g_R^2 / 3$, and $\partial 
\beta_g / \partial g^2_R = - \epsilon + (n+2) \, {\bar u}_R / 3 + n \, g_R^2$ 
at these RG $\beta$ function zeros, we find that for $\epsilon > 0$ 
($d < d_c = 4$) the {\em only} stable fixed points to one-loop order are: 
$[u^{\ast}_H , {g^\ast_C}^2]$, stable for $0< n < 4$, and
$[u^{\ast}_H , {g^\ast_0}^2]$, stable for $n \geq 4$.

The solutions of the RG equations (\ref{solcsy}) yield for the order parameter 
susceptibility and correlation function in the vicinity of an IR-stable fixed 
point the scaling laws
\begin{eqnarray}
&&\chi(\tau,{\bf q},\omega) = q^{-2-\gamma_S^\ast} \, {\hat \chi}(\tau \, 
q^{\gamma_\tau^\ast} , \omega / q^{2+a+\gamma_\lambda^\ast}) \ ,
\label{scfrm1} \\
&&C(\tau,{\bf q},\omega) = q^{-4-a-\gamma_\lambda^\ast-\gamma_S^\ast} \,
{\hat C}(\tau \, q^{\gamma_\tau^\ast} , \omega / q^{2+a+\gamma_\lambda^\ast})
\ . \nonumber \\ && \label{scfrm2}
\end{eqnarray}
Setting $\omega = 0$ in Eq.~(\ref{scfrm1}), we identify the static critical 
exponents
\begin{eqnarray}
\eta &=& -\gamma_S^\ast = 0 \ , \\
\nu^{-1} &=& -\gamma_\tau^\ast = 2 - \frac{n+2}{n+8} \ \epsilon \ ,
\end{eqnarray}
with their one-loop values computed at the Heisenberg fixed point $u_H^\ast$.
The standard hyperscaling relation then gives for the critical exponent of the
specific heat
\begin{equation}
\alpha = 2 - d \, \nu =  \frac{4-n}{2 \, (n+8)} \ \epsilon \ .
\end{equation}
Therefore, we may rewrite 
\begin{equation}
{g^\ast_C}^2 = \frac{4}{n} \ \alpha = \frac{2}{n} \ \frac{\alpha}{\nu}
\end{equation}
to this order in $\epsilon = 4-d$.
The dynamical critical exponents $z_S$ and $z_\rho$ that describe the 
divergence of the characteristic relaxation times for the order parameter and
the conserved density, respectively, are given by 
\begin{eqnarray}
z_S &=& 2 + a + \gamma_\lambda^\ast \ , \\
z_\rho &=& 2 + \gamma_D^\ast \ .
\end{eqnarray}

For model C ($a=0$) with nonconserved order parameter, we thus have {\em three}
equilibrium scaling regimes \cite{hahoma,bredom,folmos}:
(a) In the first regime with $n = 1$ (Ising symmetry), the stable critical 
one-loop fixed point is
\begin{equation}
u_H^\ast = \frac{2 \, \epsilon}{3} \ , \quad 
{g_C^\ast}^2 = \frac{2 \, \epsilon}{3} \ , \quad w_C^\ast = 1 \ .
\end{equation}
This describes a {\em strong} scaling regime where the dynamic exponents for 
the order parameter and the conserved field are identical,
\begin{equation}
z_S = z_\rho = 2 + \frac{\epsilon}{3} = 2 + \frac{\alpha}{\nu} \ .
\end{equation}
(b) In the second regime with $2 \leq n < 4$, the stable critical fixed point
becomes
\begin{equation}
u_H^\ast = \frac{6 \, \epsilon}{n+8} \ , \quad {g_C^\ast}^2 = 
\frac{2 \, (4-n)\, \epsilon}{n \, (n+8)} \ , \quad w_0^\ast = 0 \ ,
\end{equation}
leading to {\em weak} dynamic scaling with
\begin{equation}
z_S = 2 + \frac{2 \, (4-n) \, \epsilon}{n \, (n+8)} = 2 + \frac{2}{n} \ 
\frac{\alpha}{\nu} \leq z_\rho = 2 + \frac{\alpha}{\nu} \ .
\end{equation}
(c) Lastly, for $n \geq 4$,
\begin{equation}
u_H^\ast = \frac{6 \, \epsilon}{n+8} \ , \quad {g_0^\ast}^2 = 0 \ , \quad 
w_0^\ast = 0 \ , \label{modela}
\end{equation}
and consequently $z_S = z_\rho = 2$ take on their mean-field values to one-loop
order.
More generally, for $\alpha < 0$ the order parameter and conserved energy
density dynamics decouple at criticality, which implies purely model A dynamics
for the order parameter, and uncritical diffusive relaxation for the conserved
mode, i.e.,
\begin{equation}
z_S = 2 + c \, \eta \ , \quad z_\rho \equiv 2 \ ,
\end{equation}
with $c = 6 \, \ln \frac{4}{3}-1 + O(\epsilon)$.

For model D, the conserved order parameter ($a=2$) always relaxes much slower 
than the also conserved, but noncritical energy density near the phase 
transition, and consequently $w \to \infty$.
Notice that the identity $Z_\lambda \equiv Z_S$ implies 
$\gamma_\lambda \equiv \gamma_S$, and hence the model B scaling relation
$z_S \equiv 4 - \eta$ holds.
We now have only {\em two} different regimes, with the conserved field either
influenced by the critical variable, or not.
For $n < 4$ ($\alpha > 0$)
\begin{equation}
u_H^\ast = \frac{6 \, \epsilon}{n+8} \ , \ {g_C^\ast}^2 = 
\frac{2 \, (4-n)\, \epsilon}{n \, (n+8)} \ , \ w_D^\ast = \infty \ ,
\end{equation}
whence
\begin{equation}
z_S \equiv 4 - \eta \ , \quad \ z_\rho = 2 + \frac{\alpha}{\nu} \ ,
\end{equation}
whereas for $n \geq 4$ ($\alpha \leq 0$)
\begin{equation}
u_H^\ast = \frac{6 \, \epsilon}{n+8} \ , \quad {g_0^\ast}^2 = 0 \ , \quad
w_D^\ast = \infty \ ,
\end{equation}
with the decoupled model B dynamics described by
\begin{equation}
z_S \equiv 4 - \eta \ , \quad \ z_\rho \equiv 2 \ .
\end{equation}

\subsection{Isotropic detailed balance violation in model D} 
\label{isoD} 

We start with a particularly simple case of our various nonequilibrium systems,
namely that of model D subject to isotropic detailed balance violation. 
Previously, we saw upon integrating out the conserved field from the dynamic 
action and taking the asymptotic limit of the ratio of relaxation times, 
i.e., $w = D / \lambda \to \infty$, that the nonequilibrium parameter $\Theta$ 
drops out of the field theory entirely. 
This is also explicitly seen in the perturbation expansions for the vertex
functions. 
For example, consider the two point function
\begin{eqnarray*}
&&\Gamma_{\widetilde{S}S}({\bf q},0) = \lambda \, {\bf q}^2 \biggl[ r + 
\frac{n+2}{6} \, ({\widetilde u} - 3 \, {\widetilde g}^2) \int_p \! 
\frac{1}{r+{\bf p}^2} + {\bf q}^2 \nonumber \\ 
&&+ \widetilde{g}^2 (1-\Theta) \int_p \! \frac{(\frac{\bf q}{2}+{\bf p})^2}
{(\frac{\bf q}{2}+{\bf p})^2 \, [r+(\frac{\bf q}{2}+{\bf p})^2] +
w \, (\frac{\bf q}{2}-{\bf p})^2} \biggl] \, ,
\end{eqnarray*}
c.f. Eq.~(\ref{twopt1}) in appendix~\ref{appendix}.
We see that the second term which involves $\Theta$ vanishes in the limit 
$w \to \infty$. 
This is generally true of all the other vertex functions also. 
Therefore, all the effects of the nonequilibrium perturbation are eliminated.
Consequently, the renormalization factors then become identical to those of the
pure equilibrium model (\ref{zfacteq1})--(\ref{zfacteq6}), with however, the 
rescaled nonlinear couplings $u \to {\widetilde u}$ and 
$g^2 \to \widetilde{g}^2$. 
Therefore, the critical fixed points and exponents are also accordingly 
{\em identical} to those of the equilibrium model D. 
This leads us to the distinct statement that a {\em conserved} order parameter 
subject to an {\em isotropic} nonequilibrium perturbation does not display any 
novel dynamic critical behavior even when {\em quadratically} coupled to a 
conserved scalar density. 
We shall, however, see in Sec.~\ref{2TmodD} that when model D is subject to an 
{\em anisotropic} `dynamical' noise, drastic effects may emerge in this system.

\subsection{Isotropic detailed balance violation in model C}
\label{isoC} 

\subsubsection{Renormalization and one-loop RG flow functions}
For model C ($a=0$), $w < \infty$ at the stable equilibrium fixed points, so
the nonequilibrium parameter $\Theta$ does not disappear from the asymptotic
theory.
As mentioned before, a simple rescaling of the fields and coupling constants 
allows setting the relaxation rate and noise strength of the order parameter 
equal, $\lambda = \widetilde{\lambda}$ (with $k_{\rm B} T_S = 1$ here), whence 
$\Theta= \widetilde{D} / D$. 
With $Z_{\widetilde \lambda} = Z_\lambda$, the ratio of Eqs.~(\ref{zprod7}) and
(\ref{zprodIV}) gives $Z_{\widetilde S}$, and subsequently by means of
Eqs.~(\ref{zprod3}), (\ref{zprodX})--(\ref{zprodVI}) we arrive at
\begin{eqnarray}
&&Z_\rho = 1 - \widetilde{g}^2 \biggl[ \frac{n}{2} - 
\frac{2 \, (1-\Theta) \, w}{(1+w)^2} \biggr] 
\frac{A_d \mu^{-\epsilon}}{\epsilon} \ , \label{zprodisoC2} \\
&&Z_S = 1 + \widetilde{g}^2 \, \frac{(1-\Theta) \, w^2}{(1+w)^3} \
\frac{A_d \mu^{-\epsilon}}{\epsilon} \ , \label{zprodisoC4}\\
&&Z_{\widetilde S} = 1 + \frac{\widetilde{g}^2}{1+w} \label{zprodisoC5} \\ 
&&\qquad\quad\ \times \biggl[ 2 - (1-\Theta) \biggl( 1 + \frac{1}{(1+w)^2} 
\biggr) \biggr] \frac{A_d \mu^{-\epsilon}}{\epsilon} \ , \nonumber \\
&&Z_g = 1 - \frac{n+2}{3} \, \widetilde{u} \, 
\frac{A_d \mu^{-\epsilon}}{\epsilon} \label{zprodisoC6} \\
&&\quad\ + \widetilde{g}^2 \biggl[ \frac{n+4}{2} - \frac{2 \, (1-\Theta)}{1+w} 
\biggl( 1 - \frac{w}{(1+w)^2} \biggr) \biggr] 
\frac{A_d \mu^{-\epsilon}}{\epsilon} \ , \nonumber \\
&&Z_u = 1 - \frac{n+8}{6} \, \widetilde{u} \, 
\frac{A_d \mu^{-\epsilon}}{\epsilon} - \frac{6\,\widetilde{g}^4}{\widetilde{u}}
\biggl[ 1 - \frac{1-\Theta}{1+w} \biggr] \frac{A_d \mu^{-\epsilon}}{\epsilon} 
\nonumber \\
&&\quad\ + \widetilde{g}^2 \biggl[ 6 - \frac{2 \, (1-\Theta)}{1+w} \biggl( 2 + 
\frac{1}{1+w} \biggr) \biggr] \frac{A_d \mu^{-\epsilon}}{\epsilon} \ ,
\label{zprodisoC7}
\end{eqnarray}
supplementing (\ref{zprod4}), (\ref{zprod5}), (\ref{zprodI}), (\ref{zprodVII}),
and (\ref{zprodIV}).

From those renormalization constants, we infer the RG flow functions
\begin{eqnarray}
&&\gamma_S = - {\widetilde g}_R^2 \, \frac{(1-\Theta_R) \, w_R^2}{(1+w_R)^3} 
\ , \label{gammas} \\
&&\gamma_{\widetilde S} = - \frac{\widetilde{g}_R^2}{1+w_R} \biggl[ 1 + 
\Theta_R - \frac{1-\Theta_R}{(1+w_R)^2} \biggr] \ , \label{gammast} \\
&&\gamma_\lambda = \gamma_{\widetilde\lambda} = \frac{{\widetilde g}_R^2}
{1+w_R} \biggl[ 1 - \frac{1-\Theta_R}{(1+w_R)^2} \biggr] \ , \label{gammal} \\
&&\gamma_\rho = - \gamma_{\widetilde \rho} = \gamma_{\widetilde D} = 
{\widetilde g}_R^2 \biggl[ \frac{n}{2} - 
\frac{2 \, (1-\Theta_R) \, w_R}{(1+w_R)^2} \biggr] \, , \label{gammar} \\
&&\gamma_D = \frac{n}{2} \ {\widetilde g}_R^2 \ , \label{gammad} \\
&&\gamma_\tau = -2 + \frac{n+2}{6} \ \widetilde{u}_R - \widetilde{g}_R^2 
\biggl[ \frac{n+2}{2} - \frac{1-\Theta_R}{(1+w_R)^3} \biggr] \ , \nonumber 
\\ && \label{gammat}
\end{eqnarray}
and the four coupled RG $\beta$ functions
\begin{eqnarray}
&&\beta_{\widetilde u} = 6 \, \widetilde{g}_R^4 \biggl[ 1 - 
\frac{1-\Theta_R}{1+w_R} \biggr] + \widetilde{u}_R \, \biggl( -\epsilon + 
\frac{n+8}{6} \ \widetilde{u}_R \nonumber \\
&&\quad\ - 2 \, \widetilde{g}_R^2 \biggl[ 3 - \frac{1-\Theta_R}{1+w_R} \, 
\biggl( 2 + \frac{1}{1+w_R} \biggr) \biggr] \biggr) \ , \label{betau} \\
&&\beta_{\widetilde g} = \widetilde{g}_R^2 \, \biggl( -\epsilon + \frac{n+2}{3}
\ \widetilde{u}_R \label{betag} \\
&&\quad\ - 2 \, \widetilde{g}_R^2 \biggl[ \frac{n+4}{4} - 
\frac{1-\Theta_R}{1+w_R} \biggl( 1 - \frac{w_R}{(1+w_R)^2} \biggr) \biggr] 
\biggr) \ , \nonumber \\
&&\beta_w = w_R \, \widetilde{g}_R^2 \biggl[ \frac{n}{2} - \frac{1}{1+w_R} + 
\frac{1-\Theta_R}{(1+w_R)^3} \biggr] \ , \label{betaw} \\
&&\beta_\Theta = \Theta_R \, (\gamma_{\widetilde D}-\gamma_D) = - 2 \,
{\widetilde g}^2_R \, \frac{w_R \, \Theta_R \, (1-\Theta_R)}{(1+w_R)^2} \ . 
\nonumber \\ &&\label{betat} 
\end{eqnarray}

\subsubsection{RG fixed points and their stability}

For $w_R \, {\widetilde g}^2_R > 0$, Eq.~(\ref{betat}) yields three RG fixed
points for the nonequilibrium parameter $\Theta$, namely $\Theta_0^\ast = 0$, 
$\Theta_{\rm eq}^\ast = 1$, and $\Theta_\infty^\ast = \infty$.
But from
\begin{equation}
\frac{\partial\beta_\Theta}{\partial\Theta_R} = - 2 \, {\widetilde g}^2_R \,
\frac{w_R \, (1 - 2 \, \Theta_R)}{(1+w_R)^2} \label{betathp}
\end{equation}
we infer that only the {\em equilibrium} fixed point $\Theta_{\rm eq}^\ast = 1$
is stable.
This implies that detailed balance is effectively restored at the phase 
transition in this situation, and the asymptotic critical behavior is that of 
the equilibrium model C with the exponents given in Sec.~\ref{eqlbexp}.
It is however instructive to investigate the possible existence of genuine 
nonequilibrium fixed points, as they might influence the scaling behavior in 
transient crossover regimes.

We begin with $\Theta_0^\ast = 0$.
The RG $\beta$ function for the time scale ratio $w_R$ at $\Theta_R=0$ reads
\begin{equation}
\beta_w = w_R \, {\widetilde g}_R^2 \biggl[ \frac{n}{2} - \frac{w_R \, (2+w_R)}
{(1+w_R)^3} \biggr] \ , \label{th0bw}
\end{equation}
with the derivative
\begin{equation}
\frac{\partial\beta_w}{\partial w_R} = {\widetilde g}_R^2 
\biggl[ \frac{n}{2} - \frac{w_R \, (4+w_R)}{(1+w_R)^4} \biggr] \ .
\end{equation}
Since the maximum value of the second term in the brackets of Eq.~(\ref{th0bw})
at $w_R = \sqrt{3}-1$ is $\approx 0.385 < n/2$ for any $n \geq 1$, the only 
fixed points are $w_0^\ast=0$ and $w_D^\ast=\infty$, of which the former is 
stable.
At $\Theta_0^\ast = 0 = w_0^\ast$, we find 
\begin{eqnarray}
&&\gamma_S^\ast = \gamma_{\widetilde S}^\ast = \gamma_\lambda^\ast = 
\gamma_{\widetilde\lambda}^\ast = 0 \ , \\ 
&&\gamma_\rho^\ast = - \gamma_{\widetilde \rho}^\ast = \gamma_D^\ast =
\gamma_{\widetilde D}^\ast = \frac{n}{2} \ {\widetilde g}^{\ast 2} \ , \\
&&\gamma_\tau^\ast = - 2 + \frac{n+2}{6} \ {\widetilde u}^\ast - \frac{n}{2} 
\ {\widetilde g}^{\ast 2} \ ,
\end{eqnarray}
with ${\widetilde u}^\ast$ and ${\widetilde g}^{\ast 2}$ denoting the zeros of
\begin{eqnarray}
\beta_{\widetilde u} &=& {\widetilde u}_R \biggl[ - \epsilon + \frac{n+8}{6} \
{\widetilde u}_R \biggr] \ , \label{th0betu} \\
\beta_{\widetilde g} &=& {\widetilde g}_R^2 \biggl[ - \epsilon + \frac{n+2}{3}
\ {\widetilde u}_R - \frac{n}{2} \ {\widetilde g}_R^2 \biggr] \ .
\label{th0betg}
\end{eqnarray}
Note the striking similarity with the equilibrium $\beta$ functions 
(\ref{eqbetu}) and (\ref{eqbetg}), yet with the crucial sign change in 
Eq.~(\ref{th0betg}), and the fact that the anomalous dimensions at the fixed
point satisfy the relations (\ref{fdtflow}) that would be imposed by a 
fluctuation-dissipation theorem.
Upon inserting the stable (for $\epsilon > 0$) Heisenberg fixed point 
$u_H^\ast = 6 \, \epsilon / (n+8)$, we arrive at
\begin{equation}
\beta_{\widetilde g}= {\widetilde g}_R^2 \biggl[ \frac{n-4}{n+8} \ \epsilon - 
\frac{n}{2} \ {\widetilde g}_R^2 \biggr] \ ,
\end{equation}
which demonstrates that the regimes as function of $n$ for the existence of a 
nontrivial fixed point ${\widetilde g}_c^{\ast 2}$ become inverted as compared
to the equilibrium case:
${\widetilde g}_C^{\ast 2} = 2 \, (n-4) \, \epsilon / n \, (n+8) > 0$ only
for $n > 4$, but is clearly unstable.
The stable RG fixed point is thus characterized by vanishing coupling
${\widetilde g}_0^{\ast 2}=0$ to the conserved field, which again implies 
decoupled model A dynamic critical behavior, with $\eta = 0$, 
$\nu^{-1}=2-(n+2) \, \epsilon / (n+8)$, and $z_S = 2$ to one-loop order, and
purely diffusive $z_\rho \equiv 2$.
For $n < 4$, on the other hand, ${\widetilde g}_0^{\ast 2}=0$ becomes unstable.

Next, for $\Theta_\infty^\ast = \infty$, the effective dynamic coupling in 
Eqs.~(\ref{betau})--(\ref{betaw}) becomes
\begin{equation}
{\bar g}_R^2 = \Theta_R \, {\widetilde g}_R^2 \ .
\end{equation}
Thus $\gamma_D^\ast = 0$ and
\begin{equation}
\beta_w = - \, \frac{w_R \, {\bar g}_R^2}{(1 + w_R)^3} \ ,
\end{equation}
whence we see that $w_D^* = \infty$ is stable, which immediately implies that
$\gamma_S^\ast = \gamma_{\widetilde S}^\ast = 0 = \gamma_\lambda^\ast = 
\gamma_{\widetilde \lambda}^\ast$, and $\gamma_\rho^\ast = 
\gamma_{\widetilde\rho}^\ast = 0 = \gamma_{\widetilde D}^\ast$ as well.
Since $\gamma_\tau^\ast$ in Eq.~(\ref{gammat}) too reduces to the standard 
static expression, see Eq.~(\ref{gammaeq4}), this fixed point describes mere 
model A critical scaling, independent of the values of $\widetilde{u}^\ast$ and
${\bar g}^{\ast 2}$.
In fact, 
\begin{equation}
\beta_{\bar g} = {\bar g}_R^2 \biggl[ - \epsilon + \frac{n+2}{3} \ 
\widetilde{u}_R \biggr]
\end{equation}
in addition to Eq.~(\ref{th0betu}), which only allows for the standard 
decoupled model A Heisenberg fixed point $u_H^\ast = 6 \, \epsilon / (n+8)$,
${\bar g}_0^{\ast 2} = 0$ if $n \geq 4$.
As for $\Theta_0^\ast = 0$, there exists no finite nonequilibrium fixed point 
for $n < 4$.

At the unstable fixed point $\Theta_\infty^\ast = \infty$, $w_0^\ast = 0$,
\begin{eqnarray}
&&\gamma_S^\ast = 0 \, , \ \gamma_\lambda^\ast = 
\gamma_{\widetilde\lambda}^\ast = - \, \frac{\gamma_{\widetilde S}^\ast}{2} = 
{\bar g}^{\ast 2} \ , \\ 
&&\gamma_\rho^\ast = - \gamma_{\widetilde \rho}^\ast = 
\gamma_{\widetilde D}^\ast = \gamma_D^\ast = 0 \ , \\
&&\gamma_\tau^\ast = - 2 + \frac{n+2}{6} \ {\widetilde u}^\ast - 
{\bar g}^{\ast 2} \ , \\
&&\beta_{\widetilde u} = {\widetilde u}_R \biggl[ - \epsilon + \frac{n+8}{6} \
{\widetilde u}_R - 6 \, {\bar g}_R^2 \biggr] \ , \\
&&\beta_{\bar g} = {\bar g}_R^2 \biggl[ - \epsilon + \frac{n+2}{3} \ 
{\widetilde u}_R - 2 \, {\bar g}_R^2 \biggr] \ .
\end{eqnarray}
Again, the equilibrium relations (\ref{fdtflow}) are satisfied.
As to be expected, the decoupled model A fixed point 
$u_H^\ast = 6 \, \epsilon / (n+8)$, ${\bar g}_0^{\ast 2} = 0$ is stable for 
$n \geq 4$ in the $({\widetilde u}_R,{\bar g}_R^2)$ subset of parameter space, 
whereas for $n < 4$ a novel fixed point 
$u_\infty^\ast = 12 \, \epsilon / (5n+4)$, 
${\bar g}_\infty^{\ast 2} = (4-n) \, \epsilon / 2 \, (5n+4)$ emerges, with 
unusual scaling exponents $\eta = 0$, 
$\nu^{-1} = 2 - 3 \, (n+4) \, \epsilon / 2 \, (5n+4)$, 
$z_S = 2 + (4-n) \, \epsilon / 2 \, (5n+4)$, and $z_\rho = 2$ (all to one-loop 
order).
But recall that this fixed point is unstable both in the $w_R$ and the
$\Theta_R$ directions.

This leaves us with the case $w_0^* = 0$, which according to Eq.~(\ref{betat})
allows any value for the nonequilibrium parameter $\Theta_R$.
Yet since
\begin{equation}
\frac{\partial \beta_w}{\partial w_R} = {\widetilde g}_R^2 \biggl( \frac{n}{2} 
- \Theta_R \biggr)
\end{equation}
at $w_0^* = 0$, stability requires $\Theta_R \leq n/2$.
Quite generally, as in equilibrium, the model A fixed point (\ref{modela}) with
decoupled diffusive relaxation for the conserved scalar density is stable for
$n \geq 4$, with arbitrary $\Theta_R \leq n/2$, but unstable for $n < 4$.
The corresponding anomalous dimensions become
\begin{eqnarray}
&&\gamma_S^\ast = 0 \, , \ \gamma_\lambda^\ast = 
\gamma_{\widetilde\lambda}^\ast = - \, \frac{\gamma_{\widetilde S}^\ast}{2} =
{\widetilde g}^{\ast 2} \, \Theta_R \ , \\ 
&&\gamma_\rho^\ast = - \gamma_{\widetilde \rho}^\ast = \gamma_D^\ast =
\gamma_{\widetilde D}^\ast = \frac{n}{2} \ {\widetilde g}^{\ast 2} \ , \\
&&\gamma_\tau^\ast = - 2 + \frac{n+2}{6} \ {\widetilde u}^\ast - 
{\widetilde g}^{\ast 2} \biggl( \frac{n}{2} + \Theta_R  \biggr) \, .
\end{eqnarray}
Remarkably, the equilibrium relations (\ref{fdtflow}) hold once again at any 
such fixed point.
Inserting $w_0^* = 0$ into Eqs.~(\ref{betag}) and (\ref{betau}) and searching 
for nontrivial zeros leads to a quadratic equation, which is solved by
\begin{eqnarray}
&&\widetilde{g}_R^2 = \frac{\epsilon}{4 \, A} \ 
\biggl[ - 3 \, n \, (1 - 2 \, \Theta_R) \nonumber \\
&&\qquad\qquad\quad \pm \sqrt{9 \, n^2 \, (1 - 2 \, \Theta_R)^2 + 
4 \, (n-4) \, A} \ \biggr] \ , \nonumber \\
&&\textrm{and} \quad \widetilde{u}_R = \frac{3 \, \epsilon}{n+2} \biggl[ 1 + 
\frac{2 \, \widetilde{g}_R^2}{\epsilon} \, \biggl( \frac{n}{4} + \Theta_R 
\biggr) \biggr] \ , \label{quasol} \\
&&\textrm{with} \quad A = \frac{n^2 \, (n+8)}{16} + 
(5n+4) \, \Theta_R \, (1-\Theta_R) \ . \nonumber
\end{eqnarray}
\begin{figure}
\includegraphics*[scale=0.68,angle=0]{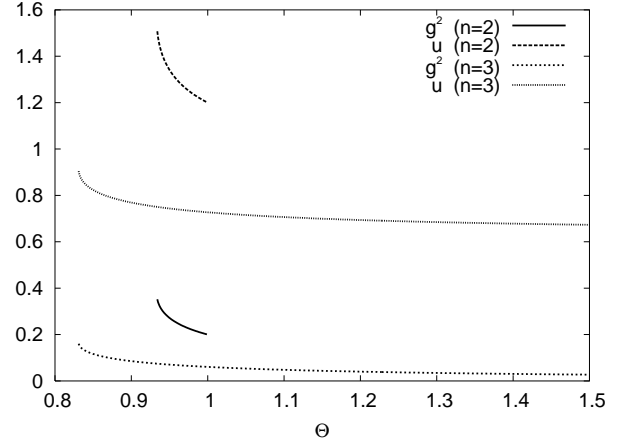}
\caption{\label{fig1} The nontrivial fixed points
   $g^2 = \widetilde{g}_R^2 / \epsilon$ and $u = \widetilde{u}_R / \epsilon$ 
   [Eq.~(\ref{quasol})] as functions of $\Theta_R \leq n/2$ for $n = 2$ 
   ($0.94 \leq \Theta_R \leq 1$) and $n = 3$ ($0.84 \leq \Theta_R \leq 1.5$).}
\end{figure}
With appropriate sign choices, this reduces to the special cases with 
$\Theta^\ast = 0$, $1$, and $\infty$ already explored above; e.g., for 
$\Theta_{\rm eq}^\ast = 1$, one finds 
$g_C^{\ast 2} = 2 \, (4-n) \, \epsilon / n \, (n+8)$ and 
$\widetilde{u}^\ast = 24 \, \epsilon / n \, (n+8)$, i.e., ${\bar u}^\ast = 
\widetilde{u}^\ast - 3 \, g_C^{\ast 2} = u_H^\ast = 6 \, \epsilon / (n+8)$.
Yet, as depicted in Fig.~\ref{fig1}, Eq.~(\ref{quasol}) permits an entire 
interval of nontrivial fixed point solutions, namely for 
$0.94 \leq \Theta_R \leq 1$ for $n=2$, and $0.84 \leq \Theta_R \leq 1.5$ for
$n=3$.
\begin{figure}[b]
\includegraphics*[scale=0.68,angle=0]{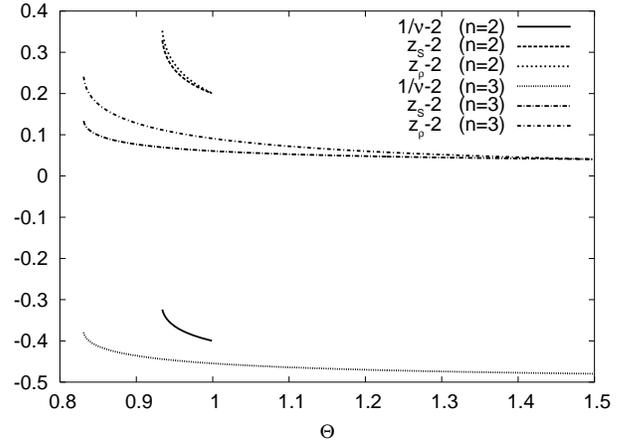}
\caption{\label{fig2} Critical exponents $\nu$, $z_S$, and $z_\rho$ for the
   isotropic nonequilibrium model C with $n = 2$ ($0.94 \leq \Theta_R \leq 1$) 
   and $n = 3$ ($0.84 \leq \Theta_R \leq 1.5$) in the weak dynamic scaling
   regime with ($w_0^\ast=0$) as functions of $\Theta_R \leq n/2$.}
\end{figure}
Apparently therefore, there exists a {\em line} of fixed points that describes 
slight perturbations from equilibrium for two- and three-component order 
parameters.
Whereas the ensuing anomalous dimensions satisfy the equilibrium constraints
(\ref{fdtflow}), the critical exponents individually differ from their
equilibrium model C values, and vary continuously as functions of the
nonequilibrium parameter $\Theta_R$, as shown in Fig.~\ref{fig2}.
We cannot exclude, however, that this unusual feature might merely constitute 
an artifact of the one-loop approximation.

In summary, for the case of isotropic detailed balance violation in model C 
with a scalar order parameter ($n = 1$), no stable genuine nonequilibrium 
fixed points are found.
The RG flow then takes the system to the {\em equilibrium} model C fixed point
with strong dynamic scaling $z_S = z_\rho = 2 + \alpha / \nu$ ($w_C^\ast = 1$),
and the standard scaling exponents as given in Sec.~\ref{eqlbexp}.
However, for model C with two- or three-component order parameter, in 
equilibrium governed by weak dynamic scaling ($z_S \leq z_\rho$, 
$w_0^\ast = 0$), at least to one-loop order lines of nonequilibrium model C 
fixed points are found that include the equilibrium fixed points, yet allow for
{\em continuously} varying static and dynamic critical exponents.
For $n \geq 4$, the conserved scalar density effectively decouples from the
order parameter, which then follows model A dynamic critical behavior.
The effective noise temperature ratio $\Theta$ naturally plays no role in this
decoupled scenario.

\section{Anisotropic violation of detailed balance in models C and D}
\label{anistp}

A spatially {\em anisotropic} nonequilibrium perturbation is applied to models
C and D by dividing our $d$-dimensional space into two sectors of 
dimensionality $d_\parallel$ and $d_\perp$ (with $d_\parallel+d_\perp=d$), and
assigning to them different noise strenghts ${\widetilde D}_\parallel$ and 
${\widetilde D}_\perp$, respectively, for the conserved energy density:
${\widetilde D} \, \nabla^2 \to {\widetilde D}_\parallel \, \nabla_\parallel^2
+ {\widetilde D}_\perp \, \nabla_\perp^2$, whence Eq.~(\ref{noiscorr2}) is 
replaced with
\begin{eqnarray}
&&\langle \eta({\bf x},t) \, \eta({\bf x}^{\prime},t^{\prime}) \rangle =
\label{aniscorr2} \\ 
&&\quad - 2 \left( {\widetilde D}_\parallel \, \nabla_\parallel^2 +
{\widetilde D}_\perp \, \nabla_\perp^2 \right) \delta({\bf x}-{\bf x}^{\prime})
\, \delta(t-t^{\prime}) \ . \nonumber
\end{eqnarray}
The conserved field noise in the two sectors can thus be thought of as being 
coupled to thermal reservoirs with different effective temperatures 
$T_\parallel$ and $T_\perp$, whence we obtain two distinct nonequilibrium 
parameters $\Theta_{\parallel / \perp} = {\widetilde D}_{\parallel / \perp} \,
\lambda / D \, {\widetilde \lambda}$.
Correspondingly, a new degree of freedom enters the problem in the form of
\begin{equation}
\sigma = \Theta_\parallel / \Theta_\perp \ , \label{sigma}
\end{equation}
the ratio of the temperatures of 
the heat baths coupled to the conserved density in the two spatial sectors, 
each measured with respect to the order parameter (critical) temperature. 
An anisotropic perturbation clearly requires that $\sigma \neq 1$.
We may choose the label assignments such that 
$\Theta_\parallel \leq \Theta_\perp$, i.e., $0 \leq \sigma \leq 1$.
In general, we must allow for the anisotropic noise in Eq.~(\ref{aniscorr2}) to
induce further splittings in the {\em renormalized} parameters $D_R$, 
$\lambda_R$, and $\widetilde{\lambda}_R$ as well (see Ref.~\cite{uwezol}).
For model D, we may in addition impose anisotropic strengths in the conserved 
order parameter noise,
\begin{eqnarray}
&&\langle \zeta^{\alpha}({\bf x},t) \, 
\zeta^{\beta}({\bf x}^{\prime},t^{\prime}) \rangle = \label{aniscorr1} \\
&&\quad - 2 \left( {\widetilde\lambda}_\parallel \, \nabla_\parallel^2 +
{\widetilde\lambda}_\perp \, \nabla_\perp^2 \right) 
\delta ({\bf x}-{\bf x}^{\prime}) \, \delta (t-t^{\prime}) \, 
\delta^{\alpha\beta} \ , \nonumber
\end{eqnarray}
whereupon $\Theta_{\parallel / \perp} = {\widetilde D}_{\parallel / \perp} \,
\lambda / D \, {\widetilde \lambda}_{\parallel / \perp}$.

\subsection{The anisotropic nonequilibrium model C}

\subsubsection{Renormalization to one-loop order}

For model C ($a=0$), we merely need to replace 
$\widetilde{D}{\bf q}^2 \to \widetilde{D}_{\parallel}{\bf q}^2_\parallel +
\widetilde{D}_{\perp}{\bf q}^2_\perp$.
Therefore, the only modifications to the previous perturbation expansions occur
in those diagrams which contain an internal conserved field propagator
$G^0_{\widetilde{\rho} \rho}({\bf q},\omega)$. 
Eqs.~(\ref{zprod4}) and (\ref{zprod6}) still hold, which implies
\begin{equation}
Z_{\rho} = Z_{\widetilde\rho}^{-1} = Z_{\widetilde{D}_{\parallel/\perp}}
\label{nc1}
\end{equation}
to all orders in perturbation theory.
Removing the logarithmic divergence in 
$\Gamma_{\widetilde{\rho}\rho}({\bf q},0)$ gives to first order
\begin{equation}
Z_D = 1 - \frac{n}{2} \ {\widetilde g}^2 \ \frac{A_d \mu^{-\epsilon}}{\epsilon}
\label{nc2}
\end{equation}
as in Eq.~(\ref{zprod5}).
The results of renormalizing those quantities that do {\em not} involve taking 
derivatives with respect to the external momenta are quite similar to the 
previous isotropic results. 
The effects of the anisotropy in these cases is a mere replacement of the
factors $1 - \Theta$ with 
$1 - (d_{\parallel} \, \Theta_{\parallel} + d_{\perp} \, \Theta_{\perp})/d\,$ 
in the expressions for the renormalization constants. 
For example, the fluctuation-induced $T_c$ shift becomes
\begin{eqnarray}
&&r_c = - \frac{n+2}{6} \, (\widetilde{u} - 3 \widetilde{g}^2) \int_p \! 
\frac{1}{r_c+{\bf p}^2} \label{anisrc} \\ 
&&\ - \frac{\widetilde{g}^2}{1+w} \left( 1 - \frac{d_\parallel}{d} \, 
\Theta_\parallel - \frac{d_\perp}{d} \, \Theta_\perp \right) \int_p \! 
\frac{1}{r_c/(1+w)+{\bf p}^2} \biggl] \, . \nonumber 
\end{eqnarray}
After rewriting in terms of the true distance from the critical point
$\tau = r - r_c$, subsequent multiplicative renormalization of 
$\Gamma_{{\widetilde S}S}(0,0)$ and 
$\partial_\omega \Gamma_{\widetilde{S}S}(0,\omega)\mid_{\omega=0}$ leads to
\begin{eqnarray}
&&(Z_{\widetilde S} \, Z_S)^{1/2} \, Z_\lambda \, Z_\tau = 1 - \biggl[ 
\frac{n+2}{6} \, (\widetilde{u} - 3 \widetilde{g}^2) \\
&&\quad + \frac{\widetilde{g}^2}{(1+w)^2} \left( 1 - \frac{d_\parallel}{d} \, 
\Theta_\parallel - \frac{d_\perp}{d} \, \Theta_\perp \right) \biggr] 
\frac{A_d \mu^{-\epsilon}}{\epsilon} \ , \nonumber \\
&&(Z_{\widetilde S} Z_S)^{1/2} = 1 + \frac{\widetilde{g}^2}{1+w} \\
&&\quad \times \biggl[ 1 - \frac{1}{1+w} \left( 1 - \frac{d_\parallel}{d} \, 
\Theta_\parallel - \frac{d_\perp}{d} \, \Theta_\perp \right) \biggr] 
\frac{A_d \mu^{-\epsilon}}{\epsilon} \ , \nonumber 
\end{eqnarray}
which are just the straightforward generalizations of Eqs.~(\ref{zprod1}) and
(\ref{zprod3}). 

Yet the resulting expressions become more complicated when the quantities to be
renormalized involve derivatives with respect to the external momentum. 
Consider 
\begin{eqnarray}
&&\frac{\partial}{\partial {\bf q}^2_{\parallel/\perp}} \, 
\Gamma_{{\widetilde S}S}({\bf q}_{\parallel/\perp},0) 
\bigg\vert_{q_{\parallel/\perp}=0} = \lambda \biggl[ 1 - 
\frac{{\widetilde g}^2}{(1+w)^2} \nonumber \\
&&\qquad\quad \times \int_p \! \biggl( 1 - 
\frac{\Theta_\parallel {\bf p}_\parallel^2 + \Theta_\perp {\bf p}_\perp^2}
{{\bf p}^2} \biggr) \frac{1}{[\tau/(1+w) + {\bf p}^2]^2} \nonumber \\
&&\quad + \frac{4 \, {\widetilde g}^2}{(1+w)^3} \ 
\frac{\partial}{\partial {\bf q}^2_{\parallel/\perp}} \\
&&\qquad\ \times \int_p \! \biggl( 1 - 
\frac{\Theta_\parallel {\bf p}_\parallel^2 + \Theta_\perp {\bf p}_\perp^2}
{{\bf p}^2} \biggr) \frac{({\bf q}_{\parallel/\perp} \cdot 
{\bf p}_{\parallel/\perp})^2}{[\tau/(1+w) + {\bf p}^2]^3} \biggr] \ ; \nonumber
\end{eqnarray}
evaluating the integrals at the normalization point $\tau=\mu^2$ in dimensional
regularization then yields
\begin{eqnarray}
&&(Z_{\widetilde S}\, Z_S)^{1/2}\, Z_{\lambda_{\parallel/\perp}} = \nonumber \\
&&\quad 1 - \frac{w \, \widetilde{g}^2}{(1+w)^3} \biggl(1 - 
\frac{d_{\parallel}}{d} \, \Theta_{\parallel} - \frac{d_{\perp}}{d} \,
\Theta_{\perp} \biggr) \frac{A_d \mu^{-\epsilon}}{\epsilon} \nonumber \\
&&\qquad \mp \frac{d_{\perp/\parallel}}{3 \, d} \ 
\frac{\widetilde{g}^2 \, (\Theta_\parallel - \Theta_\perp)}{(1+w)^3} \
\frac{A_d \mu^{-\epsilon}}{\epsilon} \ .
\end{eqnarray}

Next, from $\Gamma_{{\widetilde S}{\widetilde S}}(0,0)$, we get
\begin{equation}
Z_{\widetilde S} \, Z_{\widetilde \lambda} = 1 + \frac{\widetilde{g}^2}{1+w}
\biggl( \frac{d_{\parallel}}{d} \, \Theta_{\parallel} + \frac{d_{\perp}}{d} \, 
\Theta_{\perp} \biggr) \frac{A_d \mu^{-\epsilon}}{\epsilon} \ ,
\end{equation}
and likewise the renormalization of the three- and four-point functions results
in expressions that can simply be obtained from 
Eqs.~(\ref{zprod8})--(\ref{zprod10}) through the substitution $\Theta \to  
(d_{\parallel} \, \Theta_{\parallel} + d_{\perp} \, \Theta_{\perp}) / d$.

Upon identifying 
$Z_{\widetilde{\lambda}_{\parallel / \perp}} = Z_{\lambda_{\parallel/\perp}}$
and factoring from these products of $Z$ factors, we obtain at last
%
\begin{eqnarray}
&&Z_\rho = Z_{\widetilde\rho}^{-1} = Z_{\widetilde{D}_{\parallel/\perp}} = 1 - 
\widetilde{g}^2 \\
&&\quad \times \biggl[ \, \frac{n}{2} - \frac{2 \, w}{(1+w)^2} \biggl( 1 - 
\frac{d_{\parallel}}{d} \, \Theta_{\parallel} - \frac{d_{\perp}}{d} \, 
\Theta_{\perp} \biggr) \biggr] \frac{A_d \mu^{-\epsilon}}{\epsilon} \ , 
\nonumber \\
&&Z_{S_{\parallel/\perp}} = 1 + \frac{\widetilde{g}^2}{(1+w)^3} \, \biggl[ w^2
\biggl( 1 - \frac{d_{\parallel}}{d} \, \Theta_{\parallel} - 
\frac{d_{\perp}}{d} \, \Theta_{\perp} \biggr) \nonumber \\
&&\qquad\qquad\qquad\qquad \mp \frac{d_{\perp/\parallel}}{3 \, d} \, 
(\Theta_\parallel - \Theta_\perp) \biggr] \frac{A_d\mu^{-\epsilon}}{\epsilon} 
\ , \\
&&Z_{\widetilde{S}_{\parallel/\perp}} = 1 + \frac{\widetilde{g}^2}{1+w} \, 
\biggl[ 2 - \biggl( 1 + \frac{1}{(1+w)^2} \biggr) \\
&&\ \times \biggl( 1 - \frac{d_{\parallel}}{d} \, \Theta_{\parallel} - 
\frac{d_{\perp}}{d} \, \Theta_{\perp} \biggr) \pm 
\frac{d_{\perp/\parallel}}{3 \, d} \, 
\frac{\Theta_\parallel - \Theta_\perp}{(1+w)^2} \biggr] 
\frac{A_d \mu^{-\epsilon}}{\epsilon} \ , \nonumber \\
&&Z_{\lambda_{\parallel/\perp}} = Z_{\widetilde{\lambda}_{\parallel/\perp}} = 
1 - \frac{\widetilde{g}^2}{1+w} \, \biggl[ 1 - \frac{1}{(1+w)^2} \\
&&\ \times \biggl( 1 - \frac{d_{\parallel}}{d} \, \Theta_{\parallel} - 
\frac{d_{\perp}}{d} \, \Theta_{\perp} \biggr) \pm 
\frac{d_{\perp/\parallel}}{3 \, d} \, 
\frac{\Theta_\parallel - \Theta_\perp}{(1+w)^2} \biggr] 
\frac{A_d \mu^{-\epsilon}}{\epsilon} \ , \nonumber \\
&&Z_{\tau_{\parallel/\perp}} = 1 - \frac{n+2}{6} \, \widetilde{u} \, 
\frac{A_d \mu^{-\epsilon}}{\epsilon} \nonumber \\
&&\quad + \widetilde{g}^2 \biggl[ \, \frac{n+2}{2} - \frac{1}{(1+w)^3} 
\biggl( 1 - \frac{d_{\parallel}}{d} \, \Theta_{\parallel} - \frac{d_{\perp}}{d}
\, \Theta_{\perp} \biggr) \nonumber \\
&&\qquad\qquad\qquad\qquad \pm \frac{d_{\perp/\parallel}}{3 \, d} \ 
\frac{\Theta_\parallel - \Theta_\perp}{(1+w)^3} \biggr] 
\frac{A_d \mu^{-\epsilon}}{\epsilon} \ , \\
&&Z_{g_{\parallel/\perp}} = 1 - \frac{n+2}{3} \, \widetilde{u} \, 
\frac{A_d \mu^{-\epsilon}}{\epsilon} \nonumber \\
&&\quad + 2 \, \widetilde{g}^2 \biggl[ \, \frac{n+4}{4} - \frac{1}{1+w} 
\biggl( 1 - \frac{w}{(1+w)^2} \biggr) \\
&&\times \biggl( 1 - \frac{d_{\parallel}}{d} \, \Theta_{\parallel} - 
\frac{d_{\perp}}{d} \, \Theta_{\perp} \biggr) \pm 
\frac{d_{\perp/\parallel}}{3 \, d} \, 
\frac{\Theta_\parallel - \Theta_\perp}{(1+w)^3} \biggr] 
\frac{A_d \mu^{-\epsilon}}{\epsilon} \ , \nonumber \\
&&Z_{u_{\parallel/\perp}} = 1 - \frac{n+8}{6} \, \widetilde{u} \, 
\frac{A_d \mu^{-\epsilon}}{\epsilon} \nonumber \\
&&\quad - \frac{6 \, \widetilde{g}^4}{\widetilde{u}} \, \biggl[ 1 - 
\frac{1}{1+w} \biggl( 1 - \frac{d_{\parallel}}{d} \, \Theta_{\parallel} - 
\frac{d_{\perp}}{d} \, \Theta_{\perp} \biggr) \biggr] 
\frac{A_d \mu^{-\epsilon}}{\epsilon} \nonumber \\
&&\quad + 2 \, \widetilde{g}^2 \biggl[ 3 - \frac{1}{1+w} \biggl( 2 + 
\frac{1}{1+w} \biggr) \\
&&\times \biggl( 1 - \frac{d_{\parallel}}{d} \, \Theta_{\parallel} - 
\frac{d_{\perp}}{d} \, \Theta_{\perp} \biggr) \pm 
\frac{d_{\perp/\parallel}}{3 \, d} \, 
\frac{\Theta_\parallel - \Theta_\perp}{(1+w)^3} \biggr] 
\frac{A_d \mu^{-\epsilon}}{\epsilon} \ , \nonumber
\end{eqnarray}
in addition to Eqs.~(\ref{nc1}) and (\ref{nc2}).

\subsubsection{Anisotropic RG fixed points}

We first focus on the nonequilibrium anisotropy parameter $\sigma$.
Yet because the anisotropic contributions to the renormalization constants 
$Z_{\widetilde{\lambda}_{\parallel/\perp}}$ and $Z_{\lambda_{\parallel/\perp}}$
(even when these are not chosen identical) and 
$Z_{\widetilde{D}_{\parallel/\perp}} = Z_{D_{\parallel/\perp}}$, respectively,
are equal at least to one-loop order, its RG $\beta$ function reads
\begin{eqnarray}
\beta_\sigma &=& \sigma_R \Bigl( \gamma_{{\widetilde D}_\parallel} - 
\gamma_{D_\parallel} + \gamma_{\lambda_\parallel} - 
\gamma_{{\widetilde \lambda}_\parallel} \nonumber \\
&&\ - \gamma_{{\widetilde D}_\perp} + \gamma_{D_\perp} - 
\gamma_{\lambda_\perp} + \gamma_{{\widetilde \lambda}_\perp} \Bigr) = 0 \ ,
\end{eqnarray}
leaving the fixed point $\sigma^\ast$ undetermined.
Indeed, considering the heat bath ratios in the two sectors separately, we find
(omitting the $\parallel/\perp$ subscripts on the renormalized couplings 
$w_R$ and $\widetilde{g}^2_R$)
\begin{equation}
\beta_{\Theta_{\parallel/\perp}} = - \frac{2 \, w_R \, \widetilde{g}^2_R \, 
\Theta_{\parallel/\perp R}}{(1+w_R)^2} \left( 1 - \frac{d_\parallel}{d} \, 
\Theta_{\parallel R} - \frac{d_\perp}{d} \, \Theta_{\perp R} \right) \, , 
\label{betatpt}
\end{equation}
in almost obvious generalization of Eq.~(\ref{betat}), and the corresponding 
stability matrix becomes
\begin{eqnarray}
&&\left[ \begin{array}{cc} 
\partial \beta_{\Theta_\parallel}/\Theta_{\parallel R} &
\partial \beta_{\Theta_\parallel}/\Theta_{\perp R} \\
\partial \beta_{\Theta_\perp}/\Theta_{\parallel R}&
\partial \beta_{\Theta_\perp}/\Theta_{\perp R} \end{array} \right] = 
- \frac{2 \, w_R \, \widetilde{g}^2_R}{d \, (1+w_R)^2} \times \\
&&\ \left[ \begin{array}{cc} d - 2 \, d_\parallel \, \Theta_{\parallel R} - 
d_\perp \, \Theta_{\perp R} & - d_\perp \, \Theta_{\parallel R} \\ 
- d_\parallel \, \Theta_{\perp R} & d - d_\parallel \, \Theta_{\parallel R} - 
2 \, d_\perp \, \Theta_{\perp R} \end{array} \right] \, . \nonumber 
\end{eqnarray}

We restrict our investigation to the case $w_R \, \widetilde{g}^2_R > 0$ here,
since for weak dynamic scaling with $w_0^\ast = 0$, we already found genuine 
nonequilibrium behavior even for isotropic detailed balance violation in 
Sec.~\ref{isoC}.
We then find that the fixed points $\Theta_{\parallel/\perp 0}^\ast = 0$ and
$\Theta_{\parallel/\perp \infty}^\ast = \infty$ are unstable, whereas there is
a stable {\em line} of fixed points given by
\begin{equation}
d_\parallel \, \Theta_\parallel^\ast + d_\perp \, \Theta_\perp^\ast = d \ ,
\label{fline}
\end{equation}
which incorporates the equilibrium fixed point 
$\Theta_{\parallel {\rm eq}}^\ast = 1 = \Theta_{\perp {\rm eq}}^\ast$.
Its stability matrix eigenvalues are $0$ and $1$.
The marginal flow direction is clearly along the fixed line (arbitrary value
$\sigma^\ast$).
In fact, one readily computes for the anomalous dimensions
\begin{eqnarray}
&&\gamma_{S \parallel/\perp} = \pm \, \frac{d_{\perp/\parallel}}{3 \, d} \,
\widetilde{g}^2_R \, \frac{\Theta_\parallel-\Theta_\perp}{(1+w_R)^3} \ , \\
&&\gamma_{\widetilde{S} \parallel/\perp} = - \, 
\frac{2 \, \widetilde{g}^2_R}{1+w_R} \mp \, \frac{d_{\perp/\parallel}}{3 \, d} 
\, \widetilde{g}^2_R \, \frac{\Theta_\parallel-\Theta_\perp}{(1+w_R)^3} \ , 
\quad \\
&&\gamma_\rho = - \gamma_{\widetilde \rho} = \gamma_D = \gamma_{\widetilde D} =
\frac{n}{2} \ \widetilde{g}^2_R \ , \\
&&\gamma_{\lambda_{\parallel/\perp}}\! = 
\gamma_{\widetilde\lambda_{\parallel/\perp}}\!\! = 
\frac{\widetilde{g}^2_R}{1+w_R} \pm \frac{d_{\perp/\parallel}}{3 \, d} \, 
\widetilde{g}^2_R \frac{\Theta_\parallel-\Theta_\perp}{(1+w_R)^3} \, , \qquad\\
&&\gamma_\tau = - 2 + \frac{n+2}{6} \, {\widetilde u}_R \\
&&\qquad\qquad - \widetilde{g}^2_R \left[ \frac{n+2}{2} \pm 
\frac{d_{\perp/\parallel}}{3 \, d} \, 
\frac{\Theta_\parallel-\Theta_\perp}{(1+w_R)^3}\right] \ .
\end{eqnarray}
Consequently, the relations (\ref{fdtflow}) that follow from the 
fluctuation-dissipation theorems for the order parameter and conserved density,
respectively, are fulfilled even here.
In this sense, the entire fixed line (\ref{fline}) again represents a system 
mimicking thermal equilibrium, albeit with potentially anomalous scaling 
exponents.

Next we consider 
\begin{equation}
\beta_{w_{\parallel/\perp}} = w_R \, \widetilde{g}_R^2 \biggl[ \frac{n}{2} - 
\frac{1}{1+w_R} \mp \frac{d_{\perp/\parallel}}{3 \, d} \ 
\frac{\Theta_\parallel-\Theta_\perp}{(1+w_R)^3} \biggr] \ , \label{aniscw}
\end{equation}
with
\begin{eqnarray}
&&\frac{\partial \beta_{w_{\parallel/\perp}}}{\partial w_R} = 
{\widetilde g}_R^2 \biggl[ \frac{n}{2} - \frac{1}{(1+w_R)^2} \nonumber \\
&&\qquad\qquad \mp \frac{d_{\perp/\parallel}}{3 \, d} \, 
(\Theta_\parallel-\Theta_\perp) \, \frac{1 - 2 \, w_R}{(1+w_R)^4} \biggr] \ .
\quad
\end{eqnarray}
Thus, the weak scaling fixed point $w_0^\ast = 0$ is stable for
$n \geq 2 \pm (2 d_{\perp/\parallel}/ 3 d) \, (\Theta_\parallel-\Theta_\perp)$,
whereas $w_D^\ast = \infty$ is unstable in the $w$ direction.
In addition, there appears a {\em strong scaling} nontrivial fixed point 
$w^\ast$ given by the solution of the nonlinear equation
\begin{equation}
\frac{n}{2} \ (1+w^\ast)^3 - (1+w^\ast)^2 = 
\pm \frac{d_{\perp/\parallel}}{3 \, d} \ (\Theta_\parallel-\Theta_\perp) \ . 
\label{ancfpt}
\end{equation}
Since at this fixed point 
\begin{equation}
\frac{\partial \beta_{w_{\parallel/\perp}}}{\partial w_R} = 
\frac{2 \, w^\ast \, {\widetilde g}_R^2}{1+w^\ast} \left( \frac{3 \, n}{4} - 
\frac{1}{1+w^\ast} \right) \ ,
\end{equation}
it is stable for $n \geq 4 / 3 \, (1+w^\ast)$.
The associated anomalous dimensions become
\begin{eqnarray}
&&\gamma_S^\ast = \widetilde{g}^2_R \left( \frac{n}{2} - \frac{1}{1+w^\ast} 
\right) \ , \label{ansgas} \\
&&\gamma_\rho^\ast = \gamma_D^\ast = \gamma_\lambda^\ast = 
\frac{n}{2} \ \widetilde{g}^2_R \ , \label{ansgar} \\
&&\gamma_\tau^\ast = - 2 + \frac{n+2}{6} \, {\widetilde u}_R - 
\widetilde{g}^2_R \left( n + \frac{w^\ast}{1 + w^\ast} \right) \, . \qquad 
\label{ansgat}
\end{eqnarray}
We now need to find the RG fixed points from the coupled $\beta$ functions
\begin{eqnarray}
\beta_{\widetilde u} &=& {\widetilde u}_R \biggl[ - \epsilon + \frac{n+8}{6} \
{\widetilde u}_R - 2 \, c_u \, \widetilde{g}^2_R\biggr] + 6 \, 
\widetilde{g}^4_R \ , \qquad \label{ancbtu} \\
\beta_{\widetilde g} &=& {\widetilde g}_R^2 \biggl[ - \epsilon + \frac{n+2}{3}
\ {\widetilde u}_R - 2 \, c_g \, {\widetilde g}_R^2 \biggr] \ ,
\label{ancbtg}
\end{eqnarray}
where 
\begin{eqnarray}
c_g &=& \frac{3 n}{4} + \frac{w^\ast}{1 + w^\ast} \ , \label{deficg} \\
c_u &=& 2 + \frac{n}{2} + \frac{w^\ast}{1 + w^\ast} = c_g + 2 - \frac{n}{4} \ .
\end{eqnarray}
Note that $3 n / 4 \leq c_g \leq 1 + 3 n / 4$.
As usual, the model A fixed point $u_H^\ast = 6 \, \epsilon / (n+8)$, 
${\widetilde g}_0^{\ast 2} = 0$ is unstable for $n < 4$, but becomes stable for
$n \geq 4$.
For $n < 4$, the stable RG fixed point acquires nonzero values for both 
couplings ${\widetilde u}_R$ and ${\widetilde g}_R^2$, to be found as the
solution of the quadratic equation following from Eqs.~(\ref{ancbtu}) and 
(\ref{ancbtg}):
\begin{eqnarray}
&&{\widetilde g}_R^2 = \frac{\epsilon}{4 \, C} \left[ B \pm 
\sqrt{B^2 - 8 \, (4-n) \, C} \right] \ , \nonumber \\
&&{\widetilde u}_R = \frac{3 \, \epsilon}{n + 2} \left( 1 + 
\frac{2 \, c_g \, {\widetilde g}_R^2}{\epsilon} \right) \ , \label{anicfp} \\
&&\textrm{with} \ B = (n+2) \, (8-n) - 4 \, (4-n) \, c_g \ , \nonumber \\
&&\textrm{and} \ C = 2 \, (n+2)^2 - (n+2) \, (8-n) \, c_g + 2 \, (4-n) \, c_g^2
\, . \nonumber
\end{eqnarray}
This incorporates the equilibrium fixed point, since with $w_C^\ast = (2/n)-1$
one obtains $B = 6 \, n$ and $C = n^ 2 \, (n+8) / 8$, whence the two solutions
in Eq.~(\ref{anicfp}) reduce to the two nontrivial model C fixed points
discussed in Sec.~\ref{eqlbexp}: 
${\widetilde g}_1^{\ast 2} = 2 \, \epsilon / n$, $u_0^\ast = 0$ and
${\widetilde g}_C^{\ast 2} = 2 \, (4-n) \, \epsilon / n \, (n+8)$, 
$u_H^\ast = 6 \, \epsilon / (n+8)$ for ${\widetilde g}_R^2$ and 
${\bar u}_R = {\widetilde u}_R - 3 \, {\widetilde g}_R^2$.
\begin{figure}
\includegraphics*[scale=0.68,angle=0]{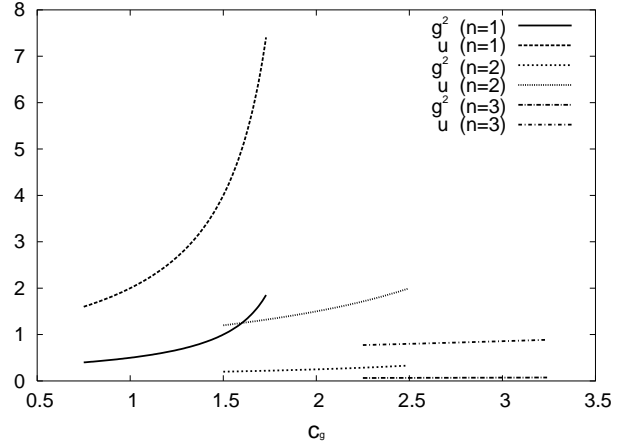}
\caption{\label{fig3} The nontrivial fixed points
   $g^2 = \widetilde{g}_R^2 / \epsilon$ and $u = \widetilde{u}_R / \epsilon$ 
   [Eq.~(\ref{anicfp})] as functions of $c_g$, 
   $3 n / 4 \leq c_g \leq 1 + 3 n / 4$ for $n=1$, $2$, and $3$.}
\end{figure}
The solutions (\ref{anicfp}) for $n=1,2,3$ are shown in Fig.~\ref{fig3} as
functions of $c_g$.
Upon inserting into the anomalous dimensions (\ref{ansgas})--(\ref{ansgat}), 
this once again yields continuously varying static and dynamic critical 
exponents $\eta = - \gamma_S^\ast$, $\nu^{-1} = - \gamma_\tau^\ast$, and
$z_S = 2 + \gamma_\rho^\ast = z_\rho$ as depicted in Figs.~\ref{fig4} and 
\ref{fig5} as functions of the parameter $c_g$.
Via Eq.~(\ref{deficg}), the parameter here is the time scale ratio $w^\ast$, or
equivalently, the effective temperature difference 
$\Theta_\parallel - \Theta_\perp$ between the longitudinal and transverse 
sectors, see Eq.~(\ref{ancfpt}). 
\begin{figure}
\includegraphics*[scale=0.68,angle=0]{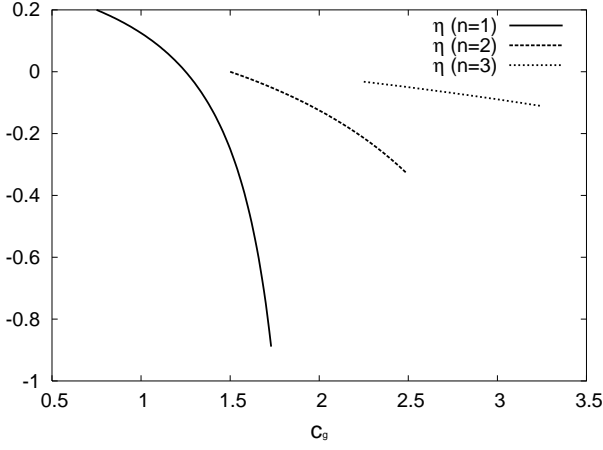}
\caption{\label{fig4} Critical exponent $\eta$ for the anisotropic 
   nonequilibrium model C with $n = 1,2,3$ as functions of the parameter $c_g$ 
   ($3 n / 4 \leq c_g \leq 1 + 3 n / 4$).}
\end{figure}
\begin{figure}[b]
\includegraphics*[scale=0.68,angle=0]{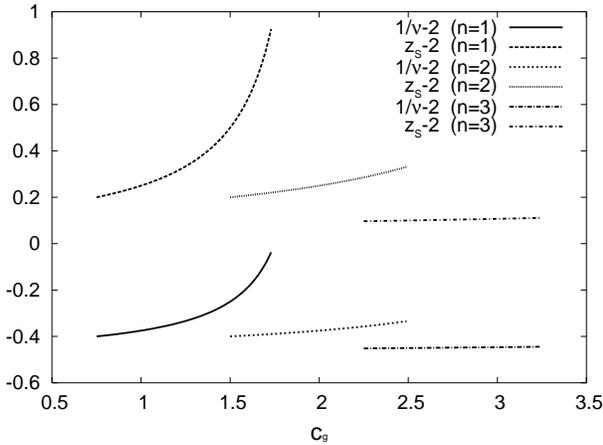}
\caption{\label{fig5} Critical exponents $\nu$ and $z_S = z_\rho$ for the
   anisotropic nonequilibrium model C with $n = 1,2,3$ as functions of $c_g$
   ($3 n / 4 \leq c_g \leq 1 + 3 n / 4$).}
\end{figure}
In conclusion, the one-loop RG flow equations for the nonequilibrium model C 
with spatially {\em anisotropic} noise allow for novel strong dynamic scaling 
regimes with $z_S = z_\rho$ with continuously varying critical exponents even 
for a scalar order parameter that encompass the equilibrium model C fixed 
point.

\subsection{The anisotropic non-equilibrium model D}

We next consider the critical behavior of our nonequilibrium version 
of model D with spatially {\em anisotropic} conserved noise.
The anisotropy may now be imposed through
$\widetilde{\lambda} \, {\bf q}^2 \to \widetilde{\lambda}_{\parallel} \,
{\bf q}^2_\parallel + \widetilde{\lambda}_{\perp} \, {\bf q}^2_\perp$ in 
addition to $\widetilde{D} \, {\bf q}^2 \to \widetilde{D}_{\parallel} \, 
{\bf q}^2_\parallel + \widetilde{D}_{\perp} \, {\bf q}^2_\perp$, see
Eqs.~(\ref{aniscorr1}) and (\ref{aniscorr2}).
We first compute the fluctuation-induced $T_c$ shift from the criticality
condition $\partial_{q^2} \, \Gamma_{\widetilde{S}S}({\bf q},0)\mid_{q=0} = 0$.
Yet for this purely relaxational dynamics, at least to one-loop order the
order parameter noise strengths $\widetilde{\lambda}_{\parallel/\perp}$ do
not enter, and we again arrive at Eq.~(\ref{anisrc}) as for model C with
nonconserved order parameter.
However, since asymptotically $w \to \infty$ here, the nonequilibrium 
parameters $\Theta_{\parallel/\perp}$ disappear, simply leaving the static 
one-loop $T_c$ shift 
\begin{equation}
r_c = - \frac{n+2}{6} \ {\bar u} \int_p \! \frac{1}{r_c+{\bf p}^2} \ .
\end{equation}
In the same manner, all other fluctuation contributions reduce to the 
equilibrium expressions.
As demonstrated explicitly in Sec.~(\ref{dftCD}) by integrating out the
conserved scalar density, the terms violating detailed balance become
obsolete at the model D fixed point $w_D^\ast = \infty$.
We remark that generalizations of dynamical models with conserved order 
parameter that contain reversible mode couplings to other conserved quantities 
behave markedly different when subject to spatially anisotropic noise
correlations:
In models H and J with `dynamical' noise, the nonlinear mode couplings induce 
{\em anisotropic} shifts of the critical temperature already to one-loop order,
thus rendering the fluctuations soft only in one subsector of momentum space.
For the ensuing two-temperature models H and J, no stable RG fixed points could
be identified, perhaps indicating that no simple nonequilibrium steady state is
approached in the long-time limit \cite{uwezol,jaiuwe,uwevjl}.

\subsection{The two-temperature model D}
\label{2TmodD}

\subsubsection{Derivation of the effective theory}

The anisotropic non-equilibrium model D discussed in the previous subsection
does not actually represent the most general spatially anisotropic extension of
relaxational dynamics with a conserved order parameter coupled to a conserved 
scalar density.
Rather, one can generalize Eqs.~(\ref{lgvneqnI}) and (\ref{lgvneqnII}) with
$a = 2$ to
\begin{eqnarray}
&&\frac{\partial S^\alpha}{\partial t} = \label{faniss} \\
&&\quad \lambda_\parallel \nabla_\parallel^2 
\biggl[ r_\parallel - \frac{\bar \lambda}{\lambda_\parallel} \, 
\nabla_\parallel^2 - 2 \, \nabla_\perp^2 + \frac{u_\parallel}{6} \sum_\beta 
{S^\beta}^2 + g_\parallel \, \rho \biggr] \, S^\alpha \nonumber \\
&&\quad + \lambda_\perp \nabla_\perp^2 \biggl[ r_\perp - \nabla_\perp^2 +
\frac{u_\perp}{6} \sum_\beta {S^\beta}^2 + g_\perp \, \rho \biggr] \, S^\alpha 
+ \zeta^\alpha \ , \nonumber \\
&&\frac{\partial \rho}{\partial t} = D_\parallel \nabla_\parallel^2 \biggl[
\rho + \frac{g_\parallel}{2} \sum_\alpha {S^\alpha}^2 \biggr] \nonumber \\
&&\qquad + D_\perp \nabla_\perp^2 \biggl[ \rho + \frac{g_\perp}{2} 
\sum_\alpha {S^\alpha}^2 \biggr] + \eta \ . \label{fanisr}
\end{eqnarray}
with noise correlators
\begin{eqnarray}
&&\langle \zeta^\alpha({\bf x},t) \, \zeta^{\beta}({\bf x}^\prime,t^\prime) 
\rangle = \label{fanins} \\
&&\qquad - 2 \, \Bigl( \widetilde{\lambda}_\parallel \, \nabla_\parallel^2 +
\widetilde{\lambda}_\perp \, \nabla_\perp^2 \Bigr) \, 
\delta ({\bf x}-{\bf x}^{\prime} \Bigr) \, \delta(t-t^{\prime}) \, 
\delta^{\alpha\beta} \ , \nonumber \\ 
&&\langle \eta({\bf x},t) \, \eta({\bf x}^{\prime},t^{\prime}) \rangle = 
\label{faninr} \\
&&\qquad - 2 \, \Bigl( \widetilde{D}_\parallel \, \nabla_\parallel^2 + 
\widetilde{D}_\perp \, \nabla_\perp^2) \, \delta ({\bf x}-{\bf x}^{\prime}) \, 
\delta (t-t^{\prime} \Bigr) \ . \nonumber 
\end{eqnarray}
We choose the labels such that $r_\perp < r_\parallel$, which is to be 
interpreted as a lower order parameter temperature $T^\perp < T^\parallel$ in 
the transverse spatial sector.
Thus, at the critical point, the longitudinal fluctuations remain uncritical 
(`stiff'), similar to equilibrium anisotropic elastic phase transitions 
\cite{elphtr}, or the behavior at Lifshitz points \cite{lifsht}.
Nonlinearities and higher-order gradient terms should then only be relevant
in the `soft' transverse directions. 
In analogy with the two-temperature nonequilibrium model B (or randomly driven
lattice gases) \cite{bearoy,kevzol,beate2}, it is possible to construct an 
effective field theory which reduces our most general anisotropic model D to an
equivalent {\em equilibrium} system, albeit with spatially long-range 
correlations. 
We first construct this effective field theory and then perform the p
erturbational renormalization of the model to one-loop order, discussing 
finally the ensuing RG flow equations.

Since $\tau_\parallel = r_\parallel - r_{c \parallel} > 0$ in the noncritical 
momentum space sector, whereas $\tau_\perp = r_\perp - r_{c \perp} \to 0$ at 
the phase transition, we expect the terms $\propto {\bf q}_\parallel^4$, 
${\bf q}_\parallel^2 \, {\bf q}_\perp^2$ to be irrelevant as compared to 
${\bf q}_\perp^4$.
In fact, in the Gaussian theory at criticality $\lambda_\parallel \, 
\tau_\parallel \, {\bf q}_\parallel^2 \sim {\bf q}_\perp^4$.
Hence we apply anisotropic scaling with $[{\bf q}_\perp] = \mu$, 
$[{\bf q}_\parallel] = [{\bf q}_\perp]^2 = \mu^2$, 
$[\omega] = [{\bf q}_\perp]^4  = \mu^4$, which yields the following scaling 
dimensions: $[{\widetilde\lambda}_\perp] = [\lambda_\perp] = \mu^0$,
$[{\widetilde\lambda}_\parallel] = [\lambda_\parallel] = \mu^{-2}$,
${\bar \lambda} = \mu^{-4}$, $[\tau_{\parallel/\perp}] = \mu^2$, 
$[{\widetilde D}_\perp] = [D_\perp] = \mu^2$, 
$[{\widetilde D}_\parallel] = [D_\parallel] = \mu^0$, and with
$[S^\alpha] = \mu^{-1+d_\parallel+d_\perp/2}$, 
$[\rho] = \mu^{d_\parallel + d_\perp/2}$ at last
$[u_{\parallel/\perp}] = [g_{\parallel/\perp}^2] = \mu^{4-d-d_\parallel}$.
Consequently, the longitudinal parameters all become {\em irrelevant} under 
scale transformations, except the marginal product 
$[\lambda_\parallel \, \tau_\parallel] = \mu^0$.
Therefore in the vicinity of the critical point, all nonlinearities in the
longitudinal sector and fluctuations $\sim {\bf q}_\parallel^4$, 
${\bf q}_\parallel^2 \, {\bf q}_\perp^2$ can be safely omitted.
From the naive scaling dimensions of $u_{\parallel/\perp}$ and 
$g_{\parallel/\perp}$ we infer the upper critical dimension 
\begin{equation}
d_c = 4 - d_\parallel \ . \label{stcrdm}
\end{equation}
It is reduced as compared to the isotropic case because the critical 
fluctuations are confined to the $d_\perp$-dimensional subsector here.

To proceed further, we rescale the fields according to 
$S^\alpha \to ({\widetilde \lambda}_\perp/\lambda_\perp)^{1/2} \, S^\alpha$, 
$\rho \to (\lambda_\perp/{\widetilde\lambda}_\perp)^{1/2} \rho$, and define
\begin{equation}
c = \frac{\lambda_\parallel}{\lambda_\perp} \, \tau_\parallel \ , \quad
\widetilde{u}_\perp = \frac{\widetilde{\lambda}_\perp}{\lambda_\perp} \ 
u_\perp \ , \quad \widetilde{g}_\perp^2 = 
\frac{\widetilde{\lambda}_\perp}{\lambda_\perp} \ g_\perp^2 \ . \label{nejcpl}
\end{equation}
The {\em effective} Langevin equations of motion for the order parameter
fields $S^\alpha$ and the conserved density $\rho$ near the phase transition
at last read
\begin{eqnarray}
\frac{\partial S^\alpha}{\partial t} &=& \lambda_\perp \left[ c \, 
\nabla_\parallel^2 + \nabla_\perp^2 \, (r_\perp - \nabla_\perp^2) \right]
S^\alpha \label{effLES} \\
&&\quad + \lambda_\perp \nabla_\perp^2 \biggl[ 
\frac{\widetilde{u}_\perp}{6} \, \sum_\beta {S^\beta}^2 + \widetilde{g}_\perp 
\, \rho \biggr] S^\alpha +\zeta^\alpha \ , \nonumber \\
\frac{\partial\rho}{\partial t} &=& D_\perp \nabla_\perp^2 \biggl[ \rho + 
\frac{\widetilde{g}_\perp}{2} \sum_\alpha {S^\alpha}^2 \biggr] \ ,
\label{effLErho}
\end{eqnarray}
with the corresponding noise correlations
\begin{eqnarray}
&&\langle \zeta^\alpha({\bf x},t) \, \zeta^\beta({\bf x}^\prime,t^\prime) 
\rangle = \nonumber \\
&&\qquad\ -2 \, \lambda_\perp \nabla_\perp^2 \, 
\delta ({\bf x}-{\bf x}^{\prime}) \, \delta(t-t^{\prime}) \, 
\delta^{\alpha\beta} \ , \label{effnscrrS} \\
&&\langle \eta({\bf x},t)\, \eta({\bf x}^\prime,t^\prime) \rangle = \nonumber\\
&&\qquad\ -2 \, D_\perp \, \Theta_\perp \, \nabla_\perp^2 \,
\delta ({\bf x}-{\bf x}^{\prime}) \, \delta (t-t^{\prime}) \ ,
\label{effnscrrrho}
\end{eqnarray}
where $\Theta_\perp$ again denotes the heat bath temperature ratio, 
\begin{equation}
\Theta_\perp = \frac{\widetilde{D}_\perp}{D_\perp} \ 
\frac{\lambda_\perp}{\widetilde{\lambda}_\perp} \ . \label{ttdneq}
\end{equation}
The preceding Eqs.~(\ref{effLES})--(\ref{effnscrrrho}) define the 
{\em two-tempera\-ture nonequilibrium model D}. 
Our analysis that led to this effective critical theory for the most general 
nonequilibrium model D with dynamical anisotropy closely parallels that of the 
two-temperature model B \cite{bearoy,kevzol,beate2}. 
Notice that after the field rescaling, only the noise strength in 
Eq.~(\ref{effnscrrrho}) violates the Einstein relation with the corresponding
relaxation constant $D_\perp$ in the critical transverse sector, if 
$\Theta_\perp \neq 1$.

Yet we can certainly write Eqs.~(\ref{effLES}) and (\ref{effLErho}) in the form
of purely relaxational Langevin dynamics
\begin{eqnarray}
\frac{\partial S^{\alpha}({\bf x},t)}{\partial t} &=& \lambda_\perp 
\nabla_\perp^2 \, 
\frac{\delta\mathcal{H}_{\rm eff}[\textbf{S},\rho]}{\delta S^\alpha({\bf x},t)}
+ \zeta^\alpha({\bf x},t) \ , \quad \\
\frac{\partial \rho({\bf x},t)}{\partial t} &=& D_\perp \nabla_\perp^2 \,
\frac{\delta \mathcal{H}_{\rm eff}[\textbf{S},\rho]}{\delta \rho({\bf x},t)} + 
\eta({\bf x},t) \ , \quad
\end{eqnarray}
with an effective Hamiltonian
\begin{eqnarray}
&&\mathcal{H}_{\rm eff}[\textbf{S},\rho] = \nonumber \\
&&\quad \int \! \frac{d^dq}{(2\pi)^d} \sum_\alpha 
\frac{c \, {\bf q}_\parallel^2 + {\bf q}_\perp^2 \, (r_\perp+{\bf q}_\perp^2)}
{2 \, {\bf q}_\perp^2} \ S^\alpha({\bf q}) \, S^\alpha(-{\bf q}) \nonumber \\
&&\quad + \int \! d^dx \, \biggl[ \frac{\widetilde{u}_\perp}{4!} 
\sum_{\alpha,\beta} S^\alpha({\bf x})^2 \, S^\beta({\bf x})^2 \nonumber \\
&&\qquad\qquad\ + \frac{\widetilde{g}_\perp}{2} \, \rho({\bf x}) \sum_\alpha 
S^\alpha({\bf x})^2 + \frac{1}{2} \, \rho({\bf x})^2 \biggr] \ . \label{effham}
\end{eqnarray}
As is obvious from its harmonic part, this Hamiltonian contains long-range 
interactions generated by the dynamical anisotropy, akin to those found in 
driven diffusive systems \cite{drdifs,bearoy}, but also at equilibrium 
ferroelastic phase transitions \cite{elphtr} and at Lifshitz points 
\cite{lifsht}.
However, our earlier investigations of model D subject to various 
nonequilibrium perturbations showed that the heat bath temperature ratio 
(\ref{ttdneq}) disappeared entirely in the asymptotic limit; since we are here 
concerned only with the transverse sector, this is reached as
$w_\perp = D_\perp / \lambda_\perp \to \infty$.
Indeed, integrating out the conserved scalar density from the dynamical action
proceeds precisely as in Sec.~\ref{dftCD}, since the dynamically generated 
long-range interactions only appear in the order parameter propagator.
As a result, the remaining detailed balance violation plays no role at all for
the fixed point properties, and the two-temperature model D in effect becomes
an {\em equilibrium} system.
Thus, we expect it to relax towards a stationary state that is characterized by
the Gibbsian probability distribution $\mathcal{P}_{\rm eq}[{\bf S},\rho] 
\propto \exp (-\mathcal{H}_{\rm eff}[{\bf S},\rho])$, with the effective 
long-range anisotropic Hamiltonian (\ref{effham}).

\subsubsection{Renormalization and critical exponents}

We introduce the renormalized fields and parameters as in Sec.~\ref{pertthry},
supplemented with
\begin{equation}
c_R = Z_c \, c \ , \quad \Theta_{\perp R} = Z_{\Theta_\perp} \, \Theta_\perp\ .
\end{equation}
But the deviation from the critical dimension now reads
\begin{equation}
\epsilon = d_c - d = 4 - d - d_\parallel = 4 - 2 \, d_\parallel - d_\perp \ ,
\end{equation}
and we define the anisotropic geometric factor as
\begin{equation}
A(d_\parallel,d_\perp) = \frac{\Gamma(3-d/2-d_\parallel/2) \, \Gamma(d/2)}
{2^{d-1} \, \pi^{d/2} \, \Gamma(d_\perp/2)} \ , \label{adpatr}
\end{equation}
with $A(0,d)=A_d$.
As a consequence of the conservation laws, and the ensuing momentum dependence
of the vertices, in analogy with the isotropic situation (see 
Sec.~\ref{pertthry}) the following relations hold to {\em all} orders in 
perturbation theory:
$\Gamma_{\widetilde{\rho}\rho}(0,\omega) \equiv i\omega$, $\partial_{q_\perp^2}
\, \Gamma_{\widetilde{\rho}\widetilde{\rho}}({\bf q}_\perp,0) \mid_{q_\perp=0} 
\equiv -2 \, D_\perp \, \Theta_\perp$, 
$\Gamma_{\widetilde{S}S}(0,\omega) \equiv i\omega$, and 
$\partial_{q_\perp^2} \, \Gamma_{\widetilde{S}\widetilde{S}}({\bf q}_\perp,0) 
\mid_{q_\perp=0} \equiv -2 \, \lambda_\perp$, whence 
$Z_{\widetilde \rho} \, Z_\rho \equiv 1$, 
$Z_{\widetilde \rho} \, Z_{D_\perp} \, Z_{\Theta_\perp} \equiv 1$,
$Z_{\widetilde S} \, Z_S \equiv 1$, and 
$Z_{\widetilde S} \, Z_{\lambda_\perp} \equiv 1$.
Notice that since the order parameter Langevin equation fulfils the Einstein
relation, this satisfies the identity 
$Z_{\lambda_\perp} = (Z_S / Z_{\widetilde S})^{1/2}$ following from the 
fluctuation-dissipation theorem
\begin{eqnarray}
\Gamma_{{\widetilde S}{\widetilde S}}({\bf q},\omega) =
\frac{2 \, \lambda_\perp \, q_\perp^2}{\omega} \ \textup{Im} \, 
\Gamma_{{\widetilde S}S}({\bf q},\omega) \ . \label{fdtverD}
\end{eqnarray}
Moreover, none of the nonlinear vertices carries transverse momentum, which
leaves the $c \, {\bf q}_\parallel^2$ term in the propagator unrenormalized to
all orders in perturbation theory as well, $\partial_{q_\parallel^2} \, 
\Gamma_{\widetilde{S}S}({\bf q}_\parallel,0) \mid_{q_\parallel=0} \equiv 
\lambda_\perp \, c$, i.e., $Z_{\lambda_\perp} \, Z_c \equiv 1$.
In summary, we obtain the exact relations
\begin{eqnarray}
&&Z_S \equiv Z_{\widetilde S}^{-1} \equiv Z_{\lambda_\perp} \equiv Z_c^{-1} \ ,
\label{zexop} \\
&&Z_\rho \equiv Z_{\widetilde \rho}^{-1} \equiv Z_{D_\perp} \, Z_{\Theta_\perp}
\ . \label{zexcd} 
\end{eqnarray}

The perturbation expansion naturally acquires the same structure as for the
equilibrium model D (or model C, with $w \to \infty$).
To one-loop order, which is determined entirely by simple combinatorics, we can
in fact immediately take over the equilibrium renormalization constants 
(\ref{zfacteq1})--(\ref{zfacteq6}) with shifted critical dimension, the
replacements $u \to \widetilde{u}_\perp$, 
$g^2 \to \widetilde{g}_\perp^2 / c^{d_\parallel/2}$, and modified geometry 
factor $A_d \to A(d_\parallel,d_\perp)$ as given in Eq.~(\ref{adpatr}).
This is confirmed explicitly by renormalizing 
\begin{eqnarray}
&&\Gamma_{{\widetilde S}S}({\bf q},0) = \lambda_\perp \biggl[ c \, 
{\bf q}_\parallel^2 + {\bf q}_\perp^4 + {\bf q}_\perp^2 \, \tau_\perp \biggl( 
1 - \frac{n+2}{6} \nonumber \\
&&\ \times (\widetilde{u}_\perp - 3 \widetilde{g}_\perp^2) \int_p 
\frac{{\bf p}_\perp^4}{[c \, {\bf p}_\parallel^2 + {\bf p}_\perp^2 
(\tau_\perp + {\bf p}_\perp)^2]^2} \biggr) \biggr]
\end{eqnarray}
at the normalization point $\tau_\perp=\mu^2$, which leads to
\begin{eqnarray}
&&Z_{\lambda_\perp} = 1 \ , \\ 
&&Z_\tau = 1 - \frac{n+2}{6} \ 
\frac{\widetilde{u}_\perp - 3 \widetilde{g}_\perp^2}{c^{d_\parallel/2}} \ 
\frac{A(d_\parallel,d_\perp) \, \mu^{-\epsilon}}{\epsilon} \ . \quad \label{zt}
\end{eqnarray}
Similarly, the logarithmic singularity in
\begin{equation}
\Gamma_{{\widetilde \rho}\rho}({\bf q},0) = D_\perp {\bf q}_\perp^2 \biggl[ 1 -
\frac{n}{2} \ \widetilde{g}_\perp^2 \int_p \frac{{\bf p}_\perp^4}{[c \, 
{\bf p}_\parallel^2 + {\bf p}_\perp^2 (\tau_\perp + {\bf p}_\perp)^2]^2}
\biggr]
\end{equation}
is absorbed into
\begin{equation}
Z_{D_\perp} = 1 - \frac{n}{2} \ \frac{\widetilde{g}_\perp^2}{c^{d_\parallel/2}}
\ \frac{A(d_\parallel,d_\perp) \, \mu^{-\epsilon}}{\epsilon} \ . \label{zD}
\end{equation}
Finally, the three- and four-point vertex functions yield
\begin{eqnarray}
&&Z_g = 1 - \frac{n+2}{3} \ \frac{\widetilde{u}_\perp}{c^{d_\parallel/2}} \ 
\frac{A(d_\parallel,d_\perp) \, \mu^{-\epsilon}}{\epsilon} \nonumber \\
&&\qquad\quad\ + \frac{n+2}{2}\ \frac{\widetilde{g}_\perp^2}{c^{d_\parallel/2}}
\ \frac{A(d_\parallel,d_\perp) \, \mu^{-\epsilon}}{\epsilon} \ ,  \label{zg} \\
&&Z_u = 1 - \frac{n+8}{6} \ \frac{\widetilde{u}_\perp}{c^{d_\parallel/2}} \ 
\frac{A(d_\parallel,d_\perp) \, \mu^{-\epsilon}}{\epsilon} \nonumber \\
&&\qquad + 6 \left( 1 - \frac{\widetilde{g}_\perp^2}{\widetilde{u}_\perp} 
\right) \frac{\widetilde{g}_\perp^2}{c^{d_\parallel/2}} \ 
\frac{A(d_\parallel,d_\perp) \, \mu^{-\epsilon}}{\epsilon} \ , \quad \label{zu}
\end{eqnarray}
as well as $Z_\rho = Z_D$, whence $Z_{\Theta_\perp} = 1$ as expected: 
Since the nonequilibrium parameter $\Theta_\perp$ disappears from the
asymptotic theory entirely, its fixed point remains undetermined, 
$\beta_{\Theta_\perp} \equiv 0$.
We also remark that the $T_c$ shift obtained from the criticality condition is
\begin{equation}
|r_{0c}| = \left[ \frac{(n+2) \, (\widetilde{u}_\perp-3 \widetilde{g}_\perp^2) 
\, A(d_\parallel,d_\perp)}{3 \, c^{d_\parallel/2} \, (d+d_\parallel-2) \, 
(4-d-d_\parallel)} \right]^{2/(4 - d - d_\parallel)} \ . \label{mdbtcs}
\end{equation}
The divergence of the denominator here indicates that in addition to the 
reduction of the upper critical dimension, the {\em lower} critical dimension 
is lowered as well to $d_{\rm lc} = 2 - d_\parallel$, just as in the
two-temperature model B \cite{bearoy,kevzol,uwezol}.

As in Sec.~\ref{RGeqns}, we can now define flow functions via logarithmic 
derivatives of the $Z$ factors with respect to the renormalization scale $\mu$,
see Eqs.(\ref{flowfct1})--(\ref{flowfct3}), with  
$\{a\} = \widetilde{u}_\perp$, $\widetilde{g}^2_\perp$, $\lambda_\perp$, $c$,
and $\tau_\perp$ here.
The solutions to the RG equations for the vertex functions are given by 
Eq.~(\ref{solcsy}), with running couplings and parameters determined by the
flow equations Eq.~(\ref{rgflow}).
The general scaling form for the renormalized order parameter response and 
correlation function thus obtained at an IR-stable fixed point becomes
\begin{eqnarray}
&&\!\!\!\chi(\tau_\perp,{\bf q}_\parallel,{\bf q}_\perp,\omega) = 
q_\perp^{-2+\eta} \, {\hat \chi}\biggl( \frac{\tau_\perp}{q_\perp^{1/\nu}},
\frac{q_\parallel}{q_\perp^{1+\Delta}},\frac{\omega}{q_\perp^z}\biggr) , 
\label{resscm} \\
&&\!\!\!C(\tau_\perp,{\bf q}_\parallel,{\bf q}_\perp,\omega) =
q_\perp^{-2-z+\eta} \, {\hat C}\biggl( \frac{\tau_\perp}{q_\perp^{1/\nu}},
\frac{q_\parallel}{q_\perp^{1+\Delta}},\frac{\omega}{q_\perp^z}\biggr) \, , 
\nonumber \\ && \label{corscm}
\end{eqnarray}
where, in addition to the usual static exponents $\eta$, $\nu$, and the dynamic
exponent $z$, the anisotropy scaling exponent $\Delta$ has been introduced.

Since the two-temperature model D is effectively in equilibrium, we may insert
the Heisenberg fixed point $u_H^\ast = 6 \, \epsilon / (n+8)$ for 
${\bar u}_\perp = \widetilde{u}_\perp - 3 \, \widetilde{g}_\perp^2$ to obtain
the static critical exponents, which thus assume the usual one-loop form
\begin{eqnarray}
&&\eta = - \gamma_S^\ast= 0 \ , \\
&&\nu^{-1} = - \gamma_{\tau_\perp}^\ast = 2 - \frac{n+2}{n+8} \ \epsilon \ , \\
&&\alpha = 2 - d \, \nu = \frac{4-n}{2 \, (n+8)} \ \epsilon \ ,
\end{eqnarray}
but with $\epsilon = 4 - d - d_\parallel$.
To two-loop order, the static critical exponents were evaluated in 
Ref.~\cite{beate2}.
In addition, upon invoking the exact relation (\ref{zexop}), i.e.,
$\gamma_S \equiv \gamma_{\lambda_\perp} \equiv - \gamma_c$, we arrive at 
\begin{eqnarray}
&&z_\rho = 2 + \gamma_{D_\perp}^\ast \ , \\
&&z_S = 4 + \gamma_{\lambda_\perp}^\ast \equiv 4 - \eta \ , \\
&&\Delta = 1 - \frac{\gamma_c^\ast}{2} \equiv 1 - \frac{\eta}{2} \ ,
\end{eqnarray}
All order parameter scaling exponents are thus given by the static critical
exponents, precisely as in the two-temperature model B 
\cite{bearoy,kevzol,beate2}.

As in equilibrium, the same is true for the dnyamic critical exponent governing
the conserved energy density $\rho$.
Since $\zeta_c = 0$ in the one-loop approximation, the ensuing RG $\beta$
functions for ${\bar u}_\perp$ and $\widetilde{g}^2_\perp$ are just 
Eqs.~(\ref{eqbetu}) and (\ref{eqbetg}) of the equilibrium model C/D.
Consequently for $n < 4$, $\widetilde{g}^2_\perp \to {g^\ast_C}^2 = 
2 \, (4-n) \, \epsilon / n \, (n+8) = 2 \, \alpha / n \, \nu$ and
\begin{equation}
z_\rho = 2 + \frac{\alpha}{\nu} \ , 
\end{equation}
whereas for $n \geq 4$, $\alpha \leq 0$ and $\widetilde{g}_\perp^2 \to 0$. 
Therefore the coupling between the order parameter and the conserved density 
becomes irrelevant, resulting in a purely diffusive 
\begin{equation}
z_\rho \equiv 2 \ .
\end{equation}
Therefore, the independent static and dynamic critical exponents to one-loop
order look identical with those of the equilibrium model D, albeit with shifted
$\epsilon = 4-d-d_\parallel$. 
The order parameter scaling exponents, including the additional anisotropy
exponent, moreover are precisely those of the two-temperature model B.

\section{Summary and conclusions}

In this paper, we have studied the critical behavior of the relaxational models
C and D with nonconserved and conserved order parameter, respectively, coupled 
to a conserved scalar density, and subject to both isotropic and anisotropic 
nonequilibrium perturbations. 
This supplements previous work on the identification of genuine nonequilibrium 
critical behavior in the form of modified dynamic universality classes in 
$O(n)$-symmetric models \cite{uwevjl}. 
These investigations have demonstrated the general robustness of the 
equilibrium critical behavior in models with nonconserved order parameter with 
respect to the violation of detailed balance, both isotropically and 
anisotropically.
This remarkable stability has been established particularly for model A which 
represents the simplest critical dynamics with a nonconserved order parameter
\cite{fhaake,grins1,kevbea}.
But even in more complicated situations involving reversible mode couplings 
between a nonconserved order parameter and additional conserved quantities, 
viz. models E, G or their $n$-component generalization, the SSS model
\cite{sssmod}, the equilibrium RG fixed point turned out to be stable, and thus
describes the asymptotic critical power laws, despite the existence of 
additional genuine nonequilibrium fixed points \cite{uwezol}.

Our results here for model C with {\em scalar} order parameter ($n = 1$), which
extends model A to include a nonlinear coupling to a conserved scalar density, 
are in accord with this general observation. 
Specifically in the case of {\em isotropic} detailed balance violation, the 
coupling of the order parameter and the conserved field to different heat baths
gives rise to the nonequilibrium parameter $\Theta$ which represents the 
temperature ratio of these heat baths. 
This variable induces different renormalizations for the noise strenghts, with 
the possibility for genuinely new dynamic as well as static critical behavior. 
Even when unstable, such nonequilibrium fixed points would affect crossover 
features and corrections to scaling in the critical regime.
However, a stability analysis yields that the equilibrium fixed point that
describes {\em strong} dynamic scaling ($w_C^\ast = 1$) with $\Theta = 1$ 
remains stable.
At least to one-loop order, we could not identify any genuine nonequilibrium 
model C scaling regime for the case of a scalar (Ising) order parameter, even
for the extreme situations with either $\Theta = 0$ or $\Theta = \infty$.
For $n=1$, the asymptotic critical behavior is thus definitely governed by the 
equilibrium model C fixed point, with the static critical exponents of the 
$O(n)$-symmetric $\phi^4$ model, and with equal dynamic exponents 
$z_S = z_\rho = 2 + \alpha / \nu$ \cite{hahoma,bredom}.
The critical behavior again reduces to that of the isotropic case as described 
above.
Therefore, we obtain the remarkable result that a quadratic coupling of a
scalar order parameter to a conserved density, which preserves the internal
symmetry of the corresponding equilibrium system, does not produce any novel 
universality classes for models with a nonconserved order parameter, subject to
detailed balance violations.
This result is to be seen in contrast with the system studied in 
Ref.~\cite{grins2} which incorpates a {\em linear} coupling of a conserved 
field to a nonconserved order parameter; in that case, effective long-range
interactions are generated, which yield novel nonequilibrium scaling features. 

For model C with $n$-component order parameter, RG fixed points with 
$\Theta^\ast \neq 1$ do appear for $n > 4$.
Yet in this situation, the order parameter effectively decouples from the 
conserved density, resulting in model A critical behavior.
However, our RG analysis yields more interesting critical scaling for the 
nonequilbrium model C with two or three order parameter components.
In equilibrium, one encounters weak dynamic scaling in these cases, with
$z_S \leq z_\rho$ ($w_0^\ast = 0$).
We find that the one-loop flow equations allow for an entire line of fixed 
points encompassing the equilibrium case.
Consequently, an interval of fixed point values emerges for the nonequilibrium 
parameter $\Theta$, leading to continuously varying static as well as dynamic 
critical exponents (as shown in Fig.~\ref{fig2}).
Curiously, the general scaling relations imposed by the fluctuation-dissipation
theorem remain satisfied along this entire fixed line.
In a similar manner, for the nonequilibrium model C with spatially 
{\em anisotropic}, dynamical noise, we obtain a line of nonequilibrium 
{\em strong} scaling fixed points for $n < 4$, i.e., even for a scalar order 
parameter, with an allowed interval of fixed point values $w^\ast$, 
characterized again by continuously varying scaling exponents 
(Figs.~\ref{fig4} and \ref{fig5}).

For the nonequilibrium model D with isotropic detailed balance violation,
the relaxation of the conserved noncritical density occurs inevitably much 
faster than that of the also conserved order parameter.
Hence the conserved energy density is always able to keep up with the critical 
fluctuations and does not in turn influence the order parameter dynamics.
This is clearly seen in the perturbation expansion upon taking the limit 
$w \to \infty$ for the diffusion rate ratio of the conserved scalar density and
order parameter, whereupon all terms involving the heat bath temperature ratio 
$\Theta$ disappear.
A further rescaling of the static nonlinearity $u \to {\widetilde u}$ and the 
coupling constant $g^2 \to {\widetilde g}^2$ then reduces this nonequilibrium 
model D variant fully to its equilibrium counterpart.

However, introducing dynamical anisotropy, i.e., different effective noise 
{\em and} ordering temperatures in the longitudinal and transverse spatial 
directions in model D with conserved order parameter has a much more drastic 
effect, since now only the momentum space sector with weaker noise softens.
As with the anisotropic nonequilibrium model B \cite{bearoy,kevzol,beate2}, it 
is possible to recast the emerging two-temperature model D with its nonlinear 
coupling to a conserved density into an effectively equilibrium model, albeit 
with a Hamiltonian that already contains long-range correlations. 
The consequences are strongly anisotropic scaling, and a reduced upper critical
dimension $d_c = 4 - d_\parallel$.
We finally remark that this feature of the two-temperature relaxational models
B and D is at variance with other conserved order parameter systems that 
incorporate {\em reversible} mode couplings to additional slow variables.
Upon introducing anisotropic dynamical noise into models J \cite{uwezol} and H 
\cite{jaiuwe}, the equilibrium integrability conditions become irretrievably 
violated; at least to one-loop order one cannot even find any stable RG fixed
points, suggesting that simple nonequilibrium steady states may not be 
sustainable in those situations.

\begin{acknowledgments}
This research has been supported through the National Science Foundation,
Division of Materials Research, grant nos. DMR 0075725 and DMR 0308548, and the
Jeffress Memorial Trust, grant no. J-594.
We gladly acknowledge helpful discussions with Jaime Santos, Beate Schmittmann,
and Royce Zia.
\end{acknowledgments}

\onecolumngrid
\appendix
\section{Explicit one-loop results for the vertex functions}
\label{appendix}

In this appendix, we list the explicit results to one-loop order in the 
perturbation expansion for the vertex functions required for the 
renormalization of the parameters and couplings of models C ($a = 0$) and D
($a = 2$). 
In the following expressions the momentum integrals are given in abbreviated 
notation, i.e., $\int_p \ldots \equiv (2\pi)^{-d} \int \! d^dp \ldots$, and the
internal frequency integrals have already been performed (via the residue 
theorem). 
We do not provide the Feynman graphs themselves, but only note the number of 
the contributing one-loop diagrams for each vertex function. 

For $\Gamma_{\widetilde{S}S}({\bf q},\omega)$, there are three one-loop graphs
that give
\begin{eqnarray}
\Gamma_{\widetilde{S}S}({\bf {\bf q}},\omega) &=& \lambda \, {\bf q}^a \Biggl[
r + \frac{n+2}{6} \ ({\widetilde u}-3{\widetilde g}^2) \int_p 
\frac{1}{r+{\bf p}^2} + {\bf q}^2 \nonumber \\
&&\qquad\quad + {\widetilde g}^2 \, (1 - \Theta) \int_p 
\frac{(\frac{\bf q}{2}+{\bf p})^a}{\frac{i\omega}{\lambda} + 
(\frac{\bf q}{2}+{\bf p})^a \, [r+(\frac{\bf q}{2}+{\bf p})^2] + 
\frac{D}{\lambda} \, (\frac{\bf q}{2}-{\bf p})^2} \Biggr] 
\nonumber \\
&&+ i\omega \Biggl[ 1 + {\bf q}^a \, {\widetilde g}^2 \int_p 
\frac{1}{r+(\frac{\bf q}{2}+{\bf p})^2} \ \frac{1}{\frac{i\omega}{\lambda}+
(\frac{\bf q}{2}+{\bf p})^a \, [r+(\frac{\bf q}{2}+{\bf p})^2] + 
\frac{D}{\lambda}(\frac{\bf q}{2}-{\bf p})^2} \Biggr] \ . \label{twopt1}
\end{eqnarray}
Only one one-loop diagram contributes to each of the other three two-point 
functions. 
The resulting expressions read
\begin{eqnarray}
\Gamma_{\widetilde{\rho}\rho}({\bf q},\omega) &=& i\omega + D \, {\bf q}^2 
\Biggl[ 1 - \frac{n}{2} \ \widetilde{g}^2 \int_p \frac{1}{r+(\frac{\bf q}{2}+
{\bf p})^2} \ \frac{1}{r+(\frac{\bf q}{2}-{\bf p})^2} \nonumber \\
&&\qquad\qquad\qquad\qquad\quad \times \Biggl( 1 - \frac{i\omega/\lambda}
{\frac{i\omega}{\lambda} + (\frac{\bf q}{2}+{\bf p})^a \, 
[r+(\frac{\bf q}{2}+{\bf p})^2] + (\frac{\bf q}{2}-{\bf p})^a \, 
[r+(\frac{\bf q}{2}-{\bf p})^2]} \Biggr) \Biggr] \ , \label{twopt2} \\
\Gamma_{\widetilde{S}\widetilde{S}}({\bf q},\omega) &=& -2 \widetilde{\lambda}
\, {\bf q}^a \Biggl [1+{\bf q}^a \, \widetilde{g}^2 \, \Theta \int_p
\frac{1}{r+(\frac{\bf q}{2}+{\bf p})^2} \ \textup{Re} \, 
\frac{1}{\frac{i\omega}{\lambda}+(\frac{\bf q}{2}+{\bf p})^a \, 
[r+(\frac{\bf q}{2}+{\bf p})^2] + \frac{D}{\lambda} \, (\frac{\bf q}{2}-
{\bf p})^2} \Biggr] \ , \label{twopt3} \\
\Gamma_{\widetilde{\rho}\widetilde{\rho}}({\bf q},\omega) &=& -2 \widetilde{D}
\, {\bf q}^2 \Biggl[ 1 + \frac{n}{2} \ {\bf q}^2 \, \frac{D}{\lambda} \,
\frac{\widetilde{g}^2}{\Theta} \int_p \frac{1}{r+(\frac{\bf q}{2}+{\bf p})^2}
\ \frac{1}{r+(\frac{\bf q}{2}-{\bf p})^2} \nonumber \\
&&\qquad\qquad\qquad\qquad\qquad\quad \times \textup{Re} \, 
\frac{1}{\frac{i\omega}{\lambda}+(\frac{\bf q}{2}+{\bf p})^a \, 
[r+(\frac{\bf q}{2}+{\bf p})^2] + (\frac{\bf q}{2}-{\bf p})^a \, 
[r+(\frac{\bf q}{2}-{\bf p})^2]} \Biggr] \ . \label{twopt4}
\end{eqnarray}

There are three one-loop diagrams that contribute to the three-point function 
$\Gamma_{\widetilde{\rho}SS}({\bf q},\omega)$.
Here, ${\bf q}$ and $\omega$ denote the wavevector and frequency of the 
outgoing external $\widetilde{\rho}$ leg.
The vertex function is evaluated at symmetric incoming labels $-{\bf q}/2$ and
$-\omega/2$ for the order parameter fields. 
Setting the external frequency $\omega$ to zero, we obtain
\begin{eqnarray}
\Gamma_{\widetilde{\rho}SS}({\bf q},0) &=& D \, {\bf q}^2 \, g \Biggl[ 1 -
\frac{n+2}{6} \ \widetilde{u} \int_p \frac{1}{r+(\frac{\bf q}{2}+{\bf p})^2}
\ \frac{1}{r+(\frac{\bf q}{2}-{\bf p})^2} \nonumber \\
&&\qquad\qquad + \widetilde{g}^2 \, \Theta \int_p \frac{(\frac{\bf q}{2}+
{\bf p})^a \, (\frac{\bf q}{2}-{\bf p})^a}{(\frac{\bf q}{2}+{\bf p})^a \,
[r+(\frac{\bf q}{2}+{\bf p})^2] + (\frac{\bf q}{2}-{\bf p})^a \, 
[r+(\frac{\bf q}{2}-{\bf p})^2]} \nonumber \\
&&\qquad\qquad\qquad\qquad \times \left( \frac{1}{(\frac{\bf q}{2}+{\bf p})^a 
\, [r+(\frac{\bf q}{2}+{\bf p})^2] + \frac{D}{\lambda} \, {\bf p}^2} + 
\frac{1}{(\frac{\bf q}{2}-{\bf p})^a \, [r+(\frac{\bf q}{2}-{\bf p})^2] +
\frac{D}{\lambda} \, {\bf p}^2} \right) \nonumber \\
&&\quad + 2 \, {\widetilde g}^2 \, \frac{D}{\lambda} \int_p 
\frac{(\frac{\bf q}{2}+{\bf p})^2}{r+{\bf p}^2} \ 
\frac{({\bf q}+{\bf p})^a}{({\bf q}+{\bf p})^a \, [r+({\bf q}+{\bf p})^2] + 
{\bf p}^a \, (r+{\bf p}^2)} \ \frac{1}{{\bf p}^a \, (r+{\bf p}^2) + 
\frac{D}{\lambda} \, (\frac{\bf q}{2}+{\bf p})^2} \Biggr] \ . \label{threept1}
\end{eqnarray}
In the same notation, four one-loop graphs yield 
\begin{eqnarray}
\Gamma_{{\widetilde S}S\rho}({\bf q},0) &=& \lambda \, {\bf q}^a \, g \Biggl[
1 - \frac{n+2}{3} \ \widetilde{u} \int_p \frac{(\frac{\bf q}{2}+{\bf p})^a}
{r+{\bf p}^2} \ \frac{1}{(\frac{\bf q}{2}+{\bf p})^a \, [r+(\frac{\bf q}{2}+
{\bf p})^2] + {\bf p}^a (r+{\bf p}^2)} \nonumber \\
&&\qquad\qquad + \widetilde{g}^2 \, \Theta \int_p \frac{(\frac{\bf q}{2}+
{\bf p})^a}{(\frac{\bf q}{2}+{\bf p})^a \, [r+(\frac{\bf q}{2}+{\bf p})^2] +
\frac{D}{\lambda} \, {\bf p}^2} \ \frac{({\bf q}+{\bf p})^a}{({\bf q}+
{\bf p})^a \, [r+({\bf q}+{\bf p})^2] + \frac{D}{\lambda} \, {\bf p}^2} 
\nonumber \\
&&\qquad\qquad + \widetilde{g}^2 \, \frac{D}{\lambda} \int_p 
\frac{({\bf q}+{\bf p})^2}{r+{\bf p}^2} \ \frac{(\frac{\bf q}{2}+{\bf p})^a}
{(\frac{\bf q}{2}+{\bf p})^a \, [r+(\frac{\bf q}{2}+{\bf p})^2] + {\bf p}^a \,
(r+{\bf p}^2)} \ \frac{1}{{\bf p}^a \, (r+{\bf p}^2) + \frac{D}{\lambda} \, 
({\bf q}+{\bf p})^2} \nonumber \\
&&\qquad\qquad + \widetilde{g}^2 \, \frac{D}{\lambda} \int_p 
\frac{(\frac{\bf q}{2}-{\bf p})^2}{r+{\bf p}^2} \ \frac{(\frac{\bf q}{2}+
{\bf p})^a}{(\frac{\bf q}{2}+{\bf p})^a \, [r+(\frac{\bf q}{2}+{\bf p})^2] + 
\frac{D}{\lambda} \, (\frac{\bf q}{2}-{\bf p})^2} \nonumber \\
&&\qquad\qquad\qquad\quad \times \left( \frac{1}{(\frac{\bf q}{2}+{\bf p})^a \,
[r+(\frac{\bf q}{2}+{\bf p})^2] + {\bf p}^a(r+{\bf p}^2)} + \frac{1}{{\bf p}^a
\, (r+{\bf p}^2) + \frac{D}{\lambda} \, (\frac{\bf q}{2}-{\bf p})^2} \right)
\Biggr] \ . \label{threept2}
\end{eqnarray}

Finally, we need the four-point vertex function 
$\Gamma_{\widetilde{S}SSS}({\bf q},0)$, for which there are ten one-loop 
Feynman diagrams.
We merely record the final result for ${\bf q} \to 0$; after a little algebra,
one arrives at
\begin{eqnarray}
&&\frac{\partial}{\partial {\bf q}^a} \, \Gamma_{\widetilde{S}SSS}({\bf q},0) 
\bigg\vert_{{\bf q} = 0} = \lambda \, u \Biggl[ 1 - \frac{n+8}{6} \ 
\widetilde{u} \int_p \frac{1}{(r+{\bf p}^2)^2} \nonumber \\ 
&&\qquad\qquad\qquad\qquad\qquad\qquad\quad + 3 \, \widetilde{g}^2 \, \Theta 
\int_p \frac{{\bf p}^a}{{\bf p}^a \, (r+{\bf p}^2) + \frac{D}{\lambda} \, 
{\bf p}^2} \left( \frac{1}{r+{\bf p}^2} + \frac{{\bf p}^a}{{\bf p}^a \, 
(r+{\bf p}^2) + \frac{D}{\lambda} \, {\bf p}^2} \right) \nonumber \\
&&\qquad\qquad\qquad\qquad\qquad\qquad\quad + 3 \, \widetilde{g}^2 
\frac{D}{\lambda} \int_p \frac{{\bf p}^2}{(r+{\bf p}^2) \, [{\bf p}^a \, 
(r+{\bf p}^2) + \frac{D}{\lambda} \, {\bf p}^2]} \left( \frac{2}{r+{\bf p}^2} +
\frac{{\bf p}^a}{{\bf p}^a \, (r+{\bf p}^2) + \frac{D}{\lambda} \, {\bf p}^2} 
\right) \nonumber \\
&&\qquad\qquad - 3 \, \frac{\widetilde{g}^4}{\widetilde{u}} \ \Theta \int_p
\frac{{\bf p}^a}{{\bf p}^a \, (r+{\bf p}^2) + \frac{D}{\lambda} \, {\bf p}^2} 
\left( \frac{1}{r+{\bf p}^2} + \frac{{\bf p}^a}{{\bf p}^a \, (r+{\bf p}^2) + 
\frac{D}{\lambda} \, {\bf p}^2} + \frac{\frac{D}{\lambda} \, {\bf p}^2}
{(r+{\bf p}^2) \, [{\bf p}^a \, (r+{\bf p}^2) + \frac{D}{\lambda} \, 
{\bf p}^2]} \right) \label{fourpt1} \\
&&\quad - 3 \, \frac{\widetilde{g}^4}{\widetilde{u}} \ \frac{D}{\lambda}
\int_p \frac{{\bf p}^2}{(r+{\bf p}^2) \, [{\bf p}^a \, (r+{\bf p}^2) + 
\frac{D}{\lambda} \, {\bf p}^2]} \left( \frac{1}{r+{\bf p}^2} + 
\frac{{\bf p}^a}{{\bf p}^a \, (r+{\bf p}^2) + \frac{D}{\lambda} \, {\bf p}^2} 
+ \frac{\frac{D}{\lambda} \, {\bf p}^2} {(r+{\bf p}^2) \, [{\bf p}^a \, 
(r+{\bf p}^2) + \frac{D}{\lambda} \, {\bf p}^2]} \right) \Biggr] \ . \nonumber
\end{eqnarray}

\twocolumngrid

\end{document}